
\input epsf
\input harvmac
\newcount\figno
\figno=0
\def\fig#1#2#3{
\par\begingroup\parindent=0pt\leftskip=1cm\rightskip=1cm\parindent=0pt
\baselineskip=11pt
\global\advance\figno by 1
\midinsert
\epsfxsize=#3
\centerline{\epsfbox{#2}}
\vskip 12pt
{\bf Fig. \the\figno:} #1\par
\endinsert\endgroup\par
}
\def\figlabel#1{\xdef#1{\the\figno}}
\def\encadremath#1{\vbox{\hrule\hbox{\vrule\kern8pt\vbox{\kern8pt
\hbox{$\displaystyle #1$}\kern8pt}
\kern8pt\vrule}\hrule}}

\overfullrule=0pt

%
\def\tilde{\widetilde}
\def\bar{\overline}
\def\np#1#2#3{Nucl. Phys. {\bf B#1} (#2) #3}
\def\pl#1#2#3{Phys. Lett. {\bf #1B} (#2) #3}

\def\prep#1#2#3{Phys. Rep. {\bf #1} (#2) #3}

\def\cmp#1#2#3{Comm. Math. Phys. {\bf #1} (#2) #3}
\def\mpl#1#2#3{Mod. Phys. Lett. {\bf #1} (#2) #3}

\font\zfont = cmss10 
\font\litfont = cmr6

\def\bigone{\hbox{1\kern -.23em {\rm l}}}
\def\ZZ{\hbox{\zfont Z\kern-.4emZ}}
\def\half{{\litfont {1 \over 2}}}

\def\CM{{\cal M}}
\def\Re{{\rm Re ~}}
\def\Im{{\rm Im ~}}
\def\lfm#1{\medskip\noindent\item{#1}}

\Title{hep-th/9407087, RU-94-52, IAS-94-43}
{\vbox{\centerline{Electric-Magnetic Duality, }
\smallskip
\centerline{Monopole Condensation, And Confinement}
\smallskip
\centerline{In $N=2$ Supersymmetric Yang-Mills Theory}}}
\smallskip
\centerline{N. Seiberg}
\smallskip
\centerline{\it Department of Physics and Astronomy}
\centerline{\it Rutgers University, Piscataway, NJ 08855-0849, USA}
\smallskip
\centerline{and}
\smallskip
\centerline{\it Institute for Advanced Study}
\centerline{\it Princeton, NJ 08540, USA}
\smallskip
\centerline{and}
\smallskip
\centerline{E. Witten}
\smallskip
\centerline{\it Institute for Advanced Study}
\centerline{\it Princeton, NJ 08540, USA}\bigskip
\baselineskip 18pt

\medskip

\noindent
We study the vacuum structure and dyon spectrum of $N=2$ supersymmetric
gauge theory in four dimensions, with gauge group $SU(2)$.  The theory
turns out to have remarkably rich and physical properties which can
nonetheless be described precisely; exact formulas can be obtained, for
instance, for electron and dyon masses and the metric on the moduli
space of vacua.  The description involves a version of Olive-Montonen
electric-magnetic duality.  The ``strongly coupled'' vacuum turns out to
be a weakly coupled theory of monopoles, and with a suitable
perturbation confinement is described by monopole condensation.
\Date{6/94}

\newsec{Introduction}

The dynamics of $N=1$ supersymmetric gauge theories in four dimensions
have been much explored, in part because of the possible
phenomenological interest. Recently results have emerged about their
strong coupling behavior
\nref\nonren{N. Seiberg, \pl{318}{1993}{469}.}%
\nref\moduli{N. Seiberg, hep-th/9402044,
Phys. Rev. {\bf D49} (1994) 6857.}%
\nref\iks{K. Intriligator, R. Leigh and N. Seiberg, hep-th/9403198,
RU-94-26, Phys. Rev. in press.}%
\nref\kl{V. Kaplunovsky and J. Louis, hep-th/9402005.}
\refs{\nonren - \kl}.  Such results are largely consequences
of the fact that the low energy
effective superpotential is holomorphic both in the chiral
superfields and in the parameters of the fundamental Lagrangian \nonren\
(other aspects of such holomorphy were discussed in
\nref\russians{M.A. Shifman and A.I. Vainshtein, \np{277}{1986}{456};
\np{359}{1991}{571}.}%
\nref\cern{D. Amati, K. Konishi, Y. Meurice, G.C. Rossi and G. Veneziano,
\prep{162}{1988}{169} and references therein.}%
\refs{\russians,\cern}).
When the behavior of the superpotential near its singularities is
combined with the symmetries and its holomorphy, the superpotential
can often be determined
exactly. The resulting superpotentials sometimes
exhibit new physical phenomena
\refs{\moduli-\kl} which might have applications also to
non-supersymmetric gauge theories.

Our goal in the present paper is, among other things, to obtain similar
strong coupling information about the corresponding $N=2$ theories, at
least in the basic case of $SU(2)$ gauge theory without matter
multiplets.  (Such multiplets will be incorporated in a subsequent paper
\ref\nnext{N. Seiberg and E. Witten, to appear.}.)
In $N=2$ theories, the Kahler potential and the masses of the stable
particles are controlled by a
holomorphic object, ``the prepotential.''  Therefore, they can be
studied in a way somewhat similar to the determination of the
superpotential in $N=1$ theories and ultimately determined.

Although our results are derived in the special case of $N=2$
supersymmetry, they exhibit
physical phenomena of general interest, including asymptotic
freedom, chiral symmetry breaking, generation of a mass scale from
strong coupling, confinement of electric charge via condensation
of magnetic monopoles, and a version of electric-magnetic duality.
These results are also likely to shed some light on the phenomena
in $N=4$ and perhaps in string theory, and are likely to help in
understanding
the topological field theories that can be obtained from $N=2$ theories
by twisting
\nref\wittentfto{E. Witten, \cmp{117}{1988}{353}.}%
\nref\wittentftt{E. Witten, IASSNS-HEP-94-5, hep-th/9403195.}%
\refs{\wittentfto,\wittentftt}.

\nref\prasad{M. K. Prasad and C. M. Sommerfield, Phys. Rev. Letters
{\bf 35} (1975) 760.}
\nref\bog{E. B. Bogomol'nyi, Sov. J. Nucl. Phys. {\bf 24} (1976) 449.}
As electric-magnetic duality will play an important role
in solving the model, we will review some of the background here.
The existence of such duality in supersymmetric Yang-Mills theories
was first suspected from properties of the
dyons -- particles carrying electric
and magnetic charge -- that exist
at the classical level in the $N=2$ and $N=4$
gauge theories.
Semiclassically one
finds -- by an argument that originated with the work of
Prasad and Sommerfeld \prasad\ and Bogomol'nyi \bog\ --
that the mass of a dyon of magnetic and electric quantum numbers
$(n_m,n_e)$ is
\eqn\howheavy{
M\geq \sqrt 2|Z|}
with
\eqn\massformula{Z =v\left(n_e+i{1\over\alpha}n_m\right).  }
Here $v$ is the Higgs expectation value, and $\alpha=g^2/4\pi$ with $g$
the gauge coupling constant.\foot{We normalize the Higgs field such that
its kinetic term is multiplied by ${1\over 4\pi \alpha}={1\over g^2}$.}
States for which the inequality in \howheavy\ is actually an equality
are said to be BPS-saturated.
To see that a similar inequality must hold quantum mechanically,
one interprets the inequality as
a consequence of a central extension of the
supersymmetry algebra
\ref\wo{E. Witten and D. Olive, \pl {78}{1978}{97}.}:
we will see later that the parameters in the central extension
(and thus the parameters appearing in the definition of $Z$) have an
interesting renormalization for $N=2$ but not for $N=4$.

The mass formula \massformula\ has a symmetry under $n_e \leftrightarrow
n_m$, $\alpha\leftrightarrow 1/\alpha$, $v\leftrightarrow v/\alpha$.
Olive and Montonen
\ref\olivemon{C. Montonen and D. Olive, \pl {72}{1977}{117}; P. Goddard,
J. Nuyts, and D. Olive, Nucl. Phys. {\bf B125} (1977) 1.}
pointed out this symmetry and conjectured that it was an exact property
of a suitable quantum theory.  According to this dramatic conjecture,
the strong coupling limit of the theory is equivalent to the weak
coupling limit with ordinary particles and solitons exchanged.  The $N=2
$ theory appears not to possess Olive-Montonen duality because the
electrons and monopoles have different Lorentz quantum numbers
(electrons are in a supersymmetric multiplet with spins $\leq 1$, while
the monopoles have spins $\leq 1/2$).  On the other hand
\ref\omfour{H. Osborn, Phys. Lett. {\bf B83} (1979) 321.},
for $N=4$ the electrons and monopoles have the same quantum numbers,
making Olive-Montonen duality more plausible.  Actually, in this paper
we will find that a version of Olive-Montonen duality holds for $N=2$.

The version in question is necessarily quite different from the one that
seems to hold for $N=4$.  The $N=2$ theory is asymptotically free, so
the coupling constant is equivalent to a choice of scale.  And an
anomaly in the $U(1)_\CR$ $R$ symmetry prevents the existence of a
physical
microscopic theta angle.  (There is still an effective theta angle at
low energies.)  For $N=4$, the anomalies cancel in both conformal
invariance and the $U(1)$ symmetry, so the theory possesses both a
natural dimensionless gauge coupling $g$ and a theta angle $\theta$; the
Olive-Montonen duality should be extended
\nref\cardy{J. Cardy and E. Rabinovici, Nucl. Phys. {\bf B205} (1982) 1;
J. Cardy, Nucl. Phys. {\bf B205} (1982) 17.}%
\nref\shap{A. Shapere and F. Wilczek, Nucl. Phys. {\bf B320}
(1989) 669.}%
to an action of
$SL(2,{\bf Z})$ on $\tau={\theta\over 2\pi} + i{4\pi\over g^2}$,
as was first recognized in lattice models \refs{\cardy,\shap}
and in string theory
\ref\ib{A. Font, L. Ibanez, D. Lust, and F. Quevedo,
Phys. Lett. {\bf B249} (1990) 35.}.
(Dramatic new evidence for this has appeared recently
\nref\sen{A. Sen, TIFR-TH-94-08, hep-th/9402002.}%
\nref\segal{G. Segal, to appear.}%
\nref\gira{L. Girardello, A. Giveon, M. Porrati and A. Zaffaroni,
NYU-TH-94/06/02, IFUM/472/FT, SISSA 80/94/EP, RI-6-94, hep-th/9406128.}%
\nref\vw{C. Vafa and E. Witten, to appear.}%
in considerations of multi-monopole bound states
\refs{\sen,\segal} and the
partition function on various manifolds \refs{\gira,\vw}.)  For $N=2$,
these natural dimensionless parameters are absent, and ``duality'' as we
interpret it has to do with the behavior of the theory as a function of
the expectation value of the Higgs field.

The organization of the paper is as follows.  Section 2 is a review of
some known facts about $N=2$ supersymmetric gauge theories.  The
classical theory has a continuous manifold of inequivalent ground states
-- the ``classical moduli space.''
Quantum corrections do not lift the vacuum degeneracy, so
also the quantum theory has a manifold of inequivalent ground states --
a ``quantum moduli space.''
We construct a low energy effective
Lagrangian for the light degrees of freedom.  As described in
\ref\natint{N. Seiberg, \pl{206}{1988}{75}.},
the metric on the quantum moduli space is
a Kahler metric that is written locally in terms of a holomorphic
function.    In certain $N=1$ models \moduli, the
quantum corrections change the topology of the moduli space.  For $N=2$
we will argue that there is no change in topology but a marked change in
geometry, singularity structure, and physical interpretation.

In section 3, we analyze more fully the geometry of the low energy
Lagrangian.  The local structure is unique only up to transformations
of a certain kind (flat space limits of the ``special geometry''
transformations
\nref\supgrat{B. De Wit and A. Van Proeyen, \np {245}{1984}89.}%
\nref\ferr{S. Ferrara,
\mpl{A6}{1991}{2175}.}%
\nref\strom{A. Strominger, Commun. Math. Phys. {\bf 133} (1990) 163.}%
\nref\candossa{P. Candelas and X. de la Ossa, Nucl. Phys. {\bf B355}
(1991) 455.}%
\refs{\supgrat-\candossa}
that appear in certain supergravity and string theories).  Physically
these correspond to duality transformations on the low energy fields.
The low energy Lagrangian is mapped to another Lagrangian of the dual
fields.

In section 4 we use the coupling of the light fields to the
massive dyons to find an expression for the dyon masses.  It includes
the quantum corrections to the classical result and is manifestly dual.
We also (following a previous two-dimensional analysis
\ref\vafa{S. Cecotti, P. Fendley, K. Intriligator, and C. Vafa,
Nucl. Phys. {\bf B386} (1992) 405; S. Cecotti and C. Vafa, Commun. Math.
Phys. {\bf 158} (1993) 569.})
describe conditions under which the spectrum of BPS-saturated
states (that is, states obeying the BPS mass formula \massformula)
does {\it not} vary continuously.  The possibility of this
phenomenon turns out to be an essential difference between $N=2$ and
$N=4$.

In sections 5 and 6 we make our proposal for solving the model. We begin
section 5 by explaining that to get a
sensible Kahler metric on the quantum moduli space, at least two
singularities are needed in strong coupling.  We propose that these
singularities arise when a magnetic monopole or dyon goes to zero mass.
As a check we show that, under a further perturbation, condensation of
monopoles occurs precisely when confinement of electric charge is
expected.  This for the first time gives a real relativistic
field theory model in
which confinement of charge is explained in this long-suspected fashion.
We also show that the monodromies resulting from massless monopoles and dyons
fit together in just the right way. Then in section 6 we show that,
with the assumption that
the singularities come from massless monopoles and dyons, it is possible
to get a unique metric on the moduli space (and unique formulas for
particle masses) obeying all the necessary conditions.

\newsec{Review of N=2 SUSY}

\subsec{Representations}

All $N=2$ theories have a global $SU(2)_R$ symmetry which acts on the
two supercharges of given chirality.  Scale invariant $N=2$ theories
have also a $U(1)_\CR $ symmetry under which the supercharges of
positive chirality have charge minus one.

We will be studying two types of $N=2$ multiplet:

\lfm{1.} The first is the $N=2$ chiral multiplet (sometimes called
a vector multiplet), containing gauge fields $A_\mu$, two Weyl fermions
$\lambda,\psi$, and a scalar $\phi$, all in the adjoint representation.
We arrange the fields as
\eqn\vecmul{\eqalign{&A_\mu \cr
\lambda ~& ~~~~\psi \cr
&\phi \cr}\quad}
to exhibit the $SU(2)_R$ symmetry which acts on the rows; $A_\mu$ and
$\phi$ are singlets and $\lambda,\psi$ are a doublet.
In terms of $N=1$ supersymmetry, these fields can be organized into
a vector multiplet $W_\alpha$ (containing $(A_\mu, \lambda)$) and
a chiral multiplet $\Phi$ (containing $(\phi,\psi)$).  In this
formalism, only one generator of $SU(2)_R$, which we will call
$U(1)_J$, is manifest.  $U(1)_J$ and $U(1)_\CR$ are both $N=1$ $R$
symmetries, acting as
\eqn\vecsym{\eqalign{
U(1)_J: \quad & \Phi \rightarrow \Phi(e^{-i\alpha} \theta) \cr
U(1)_\CR: \quad &\Phi \rightarrow e^{2i\alpha}\Phi(e^{-i\alpha} \theta)
\quad . }}

\lfm{2.} The second type of multiplet is the hypermultiplet
(sometimes called the scalar multiplet), consisting of two Weyl fermions
$\psi_q$ and $\psi^\dagger_{\tilde q}$ and complex bosons $q$ and
$\tilde q^\dagger$; $SU(2)_R$ again acts on the rows of the diamond:
\eqn\chimul{\eqalign{&\psi_q \cr
q ~& ~~~~ \tilde q^\dagger \cr
&\psi_{\tilde q}^\dagger \cr}.}
In terms of $N=1$ supersymmetry, these fields make up
two chiral multiplets $Q$ and
$\tilde Q$.  The symmetries in \vecsym\ act on them as
\eqn\chisym{\eqalign{
U(1)_J:\quad & Q\rightarrow e^{i\alpha} Q(e^{-i\alpha} \theta) \cr
& \tilde Q \rightarrow e^{i\alpha} \tilde Q(e^{-i\alpha} \theta) \cr
U(1)_\CR: \quad &Q\rightarrow  Q(e^{-i\alpha} \theta) \cr
& \tilde Q \rightarrow  \tilde Q(e^{-i\alpha} \theta) \cr}}
The gauge quantum numbers of $\tilde Q$ are dual to those of $Q$.

\subsec{Renormalizable Lagrangians and Classical Flat Directions}

We will consider an $N=2$ gauge theory, described by chiral superfields
in the adjoint representation of a gauge group $G$, possibly coupled to
additional matter hypermultiplets.  $N=2$ supersymmetry relates the
gauge couplings to certain interactions which in the $N=1$ language are
described by the superpotential.  In fact the requisite term in the
superpotential is (in the notation of
\ref\wb{J. Wess and J. Bagger, {\it Supersymmetry and Supergravity}
(Princeton University Press, 1982).})
\eqn\ntwosup{W=\sqrt{2} \tilde Q \Phi Q. }
This makes sense since $\tilde Q$ and $Q$ transform in dual
representations and $\Phi$ is in the adjoint representation.

Classically, the Lagrangian is invariant under the $SU(2)_R\times
U(1)_\CR$ symmetry \vecsym, \chisym\ as well as, possibly, some flavor
symmetries depending on the gauge representations of the hypermultiplets.
In the quantum theory, the $U(1)_\CR$ symmetry is generally broken by an
anomaly. For $SU(N_c)$ gauge symmetry with $N_f$ massless
hypermultiplets (``quarks'') in the fundamental representation,
$U(1)_\CR$ is broken to ${\bf Z}_{4N_c-2N_f}$.  In this paper, we will
mainly consider the $SU(2)$ gauge theory without quarks.  In this case
the global symmetry is $(SU(2)_R \times {\bf Z}_8)/{\bf Z}_2$ since
$4N_c-2N_f=8$ (the division by ${\bf Z}_2$ arises because the center of
$SU(2)_R$ is contained in ${\bf Z}_8$).

The classical potential of the pure $N=2$ theory (without hypermultiplets)
is
\eqn\claspot{V(\phi) = {1 \over g^2 }  \Tr [\phi,\phi^\dagger]^2.}
For this to vanish, it is not necessary that $\phi$ should vanish; it is
enough that $\phi$ and $\phi^\dagger$ commute.  The classical theory
therefore has a family of vacuum states.  For instance, if the gauge
group is $SU(2)$, then up to gauge transformation we can take $\phi= {1
\over  2} a \sigma^3$, with $\sigma^3={\rm diag}(1,-1)$ and $a$ a
complex parameter labeling the vacua.  The Weyl group of $SU(2)$ acts by
$a\leftrightarrow -a$, so the gauge invariant quantity parametrizing the
space of vacua is $u=\half a^2= \Tr \,\phi^2$.  For non-zero $a$ the gauge
symmetry is broken to $U(1)$ and the global ${\bf Z}_8$ symmetry is
broken to ${\bf Z}_4$.  The residual ${\bf Z}_4$ acts trivially on the
$u$ plane since the $U(1)_\CR$ charge of $u$ is 4.  The global symmetry
group acts on the $u$ plane as a spontaneously broken ${\bf Z}_2$,
acting by $u\leftrightarrow -u$.

Classically, there is a singularity at $u=0$, where the full $SU(2)$
gauge symmetry is restored and more fields become massless.

\subsec{Low Energy Effective Action}

We now study the low energy effective action of the light fields on the
moduli space.  For generic $\langle\phi\rangle$ the low energy effective
Lagrangian contains a single $N=2$ vector multiplet, $\CA$.  The terms
with at most two derivatives and not more than four fermions are
constrained by $N=2$ supersymmetry.  They are expressed in terms of a
single holomorphic function $\CF(\CA)$, as explained in \natint.  In
$N=1$ superspace, the Lagrangian is
\eqn\ntweffl{{1 \over 4\pi} \Im \left[ \int d^4 \theta
{\partial \CF(A)\over\partial A}
\bar A + \int d^2\theta \half  {\partial^2 \CF(A)\over\partial A^2}
W_\alpha W^\alpha \right]  }
where $A$ is the $N=1$ chiral multiplet in the $N=2$ vector multiplet
$\CA$ whose scalar component is $a$.

We would like to make a few comments:

\lfm{1.}    For large $a$, asymptotic freedom takes
over and the theory is weakly coupled.  Moreover, since it is impossible
to add an $N=2$ invariant superpotential to \ntweffl, the vacuum
degeneracy cannot be removed quantum mechanically.  Therefore, the
quantum theory has a non-trivial moduli space which is in fact a one
complex dimensional Kahler manifold.  The Kahler potential can be
written in terms of the effective low energy $\CF$ function as
\eqn\kpot{K=\Im \left({\partial \CF(A) \over \partial A} \bar A\right).}
The metric is thus concretely
\eqn\metm{(ds)^2= \Im {\partial^2 \CF(a)\over \partial a^2} \,\,\,\,
 da\,d\bar a.}
In the classical theory, $\CF$ can be read off from the tree level
Lagrangian of the $SU(2)$ gauge theory and is
$\CF(\CA)=\half \tau_{cl}A^2$ with $\tau_{cl}={\theta \over 2\pi} +
i{4 \pi \over g^2}$.  Asymptotic
freedom means that this formula is valid for large $a$ if $g^2$ is
replaced
by a suitable effective coupling.  The small $a$ behavior will however
turn out to be completely different.  Classically, the $\theta$ parameter
has no consequences.  Quantum mechanically, the physics is
$\theta$ dependent,
but since there is an anomalous symmetry, it can be absorbed in a
redefinition of the fields.  Therefore, we will set $\theta=0$.

\lfm{2.} The formula for the Kahler potential does not look covariant --
the Kahler potential can be written in this way only in a distinguished
class of coordinate systems, which we will analyze later.  In fact, $A$
is related by $N=2$ supersymmetry to the ``photon'' $A_\mu$, which has a
natural linear structure; this gives a natural coordinate system (or
what will turn out to be a natural class of coordinate systems) for $A$.

\lfm{3.} The low energy values of the gauge coupling constant and theta
parameter can be read off from the Lagrangian.  If we combine them in
the form $\tau ={\theta \over 2\pi} + i {4\pi \over g^2}$, and denote
the effective couplings in the vacuum parametrized by $a$ as $\tau(a)$,
then
\eqn\lowenga{\tau (a)= {\partial^2 \CF\over \partial a^2}.}

\lfm{4.} The generalization to an arbitrary compact gauge group $G$ of
rank $r$ is as follows.  The potential is always given by \claspot, so
the classical vacua are labeled by a complex adjoint-valued matrix
$\phi$ with $[\phi,\phi^\dagger]=0$.  The unbroken gauge symmetry at the
generic point on the moduli space is the Cartan subalgebra and therefore
the complex dimension of the moduli space is $r$.  The low energy
theory is described in terms of $r$ abelian chiral multiplets $\CA^i$,
and the generalization of \ntweffl\ is \natint
\eqn\ntwefflon{{1 \over 4\pi} \Im \left[ \int d^4 \theta {\partial
\CF(A)\over\partial A^i}
\bar A^i + \int d^2\theta \half{ \partial^2
\CF(A)\over\partial A^i\partial A^j} W_\alpha^i W^{\alpha j} \right].  }
Here $i$ labels the generators in the Cartan subalgebra and locally
${\cal F}$ is an arbitrary holomorphic function of $r$ complex variables.

\lfm{5.} The $SU(2)$ theory, studied on the flat
direction with $u\not= 0 $, has in addition to the massless chiral or
vector multiplet $\CA$, additional charged massive vector multiplets.
One can easily write a gauge invariant effective action for the triplet
of chiral multiplets $\CA^a,\,\, a=1\dots 3$, which reduces at low
energies to \ntweffl\ for the massless fields and incorporates the
massive ones.  Using the same function $\CF$ as above, we set
$\CF(\sqrt{\CA\cdot \CA}) =\CH(\CA\cdot\CA)$ and write
\eqn\ntweffln{{1 \over 2\pi} \Im \left[ \int d^4 \theta \CH^\prime A^a
\left( e^V \right)_{ab} \bar A^b + \int d^2\theta \half \left(
\CH^\prime \delta^{ab} + 2 \CH^{\prime \prime} A^a A^b \right)
W_\alpha^a W_\alpha^b \right]  }
where we used the $SU(2)$-invariant
metric $\delta^{ab}$ to raise and lower indices.
\ntweffln\ has $N=2$ supersymmetry and manifest gauge invariance,
and reduces at low energies to \ntweffl.

\lfm{6.} The Lagrangian \ntweffl\ is unchanged if we add to $\CF$ terms
linear in $\CA$.  This has the effect of shifting $\partial \CF/\partial
A$ by a constant.  We will later assign physical meaning to
\eqn\defh{h(A)={\partial \CF\over\partial A} .}
The additive constant will always be fixed by comparing with the
high energy theory as in \ntweffln.

As we have already mentioned, classically the $\CF$ function
is
\eqn\clasval{\CF_0=\half\tau_{cl} \CA^2.}
The quantum corrections
were analyzed in \natint.
The tree level and one loop contributions add up to
\eqn\cfone{\CF_{one~loop} =i {1\over 2\pi} \CA^2\ln {\CA^2 \over
\Lambda^2}}
where $\Lambda$ is the dynamically generated scale.  This logarithm is
related to the one loop beta function and also ensures the anomalous
transformation laws under $U(1)_\CR$.  Higher order perturbative
corrections are absent.  Instantons lead to new terms.  The anomaly and
the instanton action suggest that
\eqn\cfins{\CF= i {1\over 2\pi} \CA^2\ln {\CA^2 \over \Lambda^2} +
\sum_{k=1}^\infty \CF_k \left( {\Lambda \over \CA} \right)^{4k} \CA^2 }
where the $k$'th term arises as a contribution of $k$ instantons.
A detailed  calculation of the $k=1$ term \natint\ indicates that $\CF_1
\not= 0$.  We will soon see that infinitely many $\CF_k$ are nonzero.

Corrections to the classical formula \clasval\ are related to
the beta function, and
for $N=4$ supersymmetric Yang-Mills theory, whose beta function vanishes,
the formula \clasval\ is exact.

\newsec{Duality}

We have noted above that locally, by virtue of $N=2$
supersymmetry, the metric on the moduli
space is of the form
\eqn\locform{ (ds)^2= \Im\, \tau(a) \,\,da\,d\bar a,      }
with $\tau (a)$ the holomorphic function $\tau=\partial^2\CF/\partial
a^2$. The one loop formula \cfone\ shows
that for large $|a|$, $\tau (a) \approx i\left(\ln(a^2/\Lambda^2)+ 3
\right)/\pi$
is a multivalued function whose imaginary part is single-valued and positive.
However, if $\Im \tau(a)$ is globally defined it cannot be positive
definite as the harmonic function $\Im \tau $ cannot have a minimum.
This indicates that the above description of the metric must be valid
only locally.

To what extent is it possible to change variables from $a$ to some other
local parameter, while leaving the metric in the form \locform? The
answer to this question is at the heart of the physics.
We define $a_D=\partial \CF/\partial a$.
The metric can then be written
\eqn\symmet{(ds)^2=\Im da_D \,d\bar a=-{i\over 2}\left(da_D d\bar a-da
d\bar a_D\right) .  }
This formula is completely symmetric in $a$ and $a_D$, so if we use
$a_D$ as the local parameter, the metric will be in the same general
form as \locform, with a different harmonic
function replacing $\Im \tau$.  As we will see presently, this
transformation corresponds to electric-magnetic duality.
Before entering into that, let us identify the complete class of local
parameters in which the metric can be written as in \symmet.

\subsec{Mathematical Description}

To treat the formalism in a way that is completely symmetric between $a$
and $a_D$, we introduce an arbitrary local holomorphic coordinate $u$,
and treat $a$ and $a_D$ as functions of $u$.  $u$ is a local coordinate
on a complex manifold $\CM$ -- the moduli space of vacua of the theory.
Eventually we will pick $u$ to be the expectation value of
$\Tr\,\phi^2$ --
a good physical parameter -- but for now $u$ is arbitrary.

Introduce a two dimensional complex space $X\cong {\bf C}^2$ with
coordinates $(a_D,a)$.  Endow $X$ with the symplectic form $\omega= \Im
da_D\wedge d\bar a $.  The functions $(a_D(u),a(u))$ give a map $f$ from
$\CM$ to $X$.  The metric on $\CM$ is
\eqn\genform{
(ds)^2 = \Im {da_D\over du}{d\bar a\over d\bar u} du d\bar u = - { i
\over 2} \left({da_D\over du}{d\bar a\over d\bar u}- {da\over du}{d\bar
a_D \over d\bar u}\right)  du d\bar u. }
This formula is valid for an arbitrary local parameter $u$ on $\CM$.  If
one picks $u=a$, one gets back the original formula \metm\ for the
metric. (The above formula can be described in a coordinate-free way by saying
that the Kahler form of the induced metric on $\CM$ is $f^*(\omega)$.)
Notice that $\omega$ had no particular positivity property and thus, if
$a(u)$ and $a_D(u)$ are completely arbitrary local holomorphic
functions, the metric \genform\ is not positive.  We will eventually
construct $a(u)$ and $a_D(u)$ in a particular way that will ensure
positivity.

It is easy to see what sort of transformations preserve the general
structure of the metric.  If we set $a^\alpha=(a_D,a),\,\,\alpha=1,2$,
and let $\epsilon_{\alpha\beta}$ be the antisymmetric tensor with
$\epsilon_{12}=1$, then
\eqn\newtry{(ds)^2=-{i\over 2}\epsilon_{\alpha\beta} {da^\alpha\over du}
{ d\bar a^\beta\over d\bar u} du d\bar u .}
This is manifestly invariant under linear transformations that
preserve $\epsilon$ and commute with complex conjugation (the latter
condition ensures that $a^\alpha$ and $\bar a^\alpha$ transform the same
way).  These transformations make the group $SL(2,{\bf R})$
(or equivalently $Sp(2,{\bf R})$).  Also, \newtry\ is obviously
invariant under adding a constant to $a_D$ or $a$.  So if
we arrange $(a_D,a)$ as a column vector
$v$, the symmetries that preserve the general structure are
\eqn\transfor{v \to Mv + c,}
where $M$ is a $2\times 2$ matrix in $SL(2,{\bf R})$, and $c$ is a
constant vector.  Later we will find (from considerations involving the
gauge fields and the electric and magnetic charges) that $M$ must be in
$SL(2,{\bf Z})$ and that in the
pure $N=2$
gauge theory, $c$ must vanish.  (In coupling to matter, $c$ will play
an important role \nnext.)  In general, the group of transformations
\transfor\ can be thought of as the group of $3\times 3$ matrices of the
form
\eqn\specmat{\left(\matrix{ 1 & 0 \cr
                            c & M \cr}\right),}
acting on the three objects $(1,a_D,a)$.

{\it {Generalization To Dimension Greater Than One}}

Though it will not be exploited in the present paper, and can thus
be omitted by the reader, let us
briefly discuss the generalization to other gauge groups.
If the gauge group $G$ has rank $r$, then
$\CM$ has complex dimension $r$.  Locally from \ntwefflon\ the metric
is
\eqn\manloc{(ds)^2=\Im {\partial^2\CF\over\partial a^i\partial
a^j}da^id\bar a^j ,}
with distinguished local coordinates $a^i$ and a holomorphic function
$\CF$.  We again reformulate this by introducing
\eqn\dualvar{
a_{D,j}={\partial \CF\over\partial a^j}.}
Then we can write
\eqn\canwr{(ds)^2=\Im \sum_i  da_{D,i}\, d\bar a^i .}
To formulate this invariantly, we introduce a complex space $X\cong {\bf
C}^{2r} $ with coordinates $a^i,a_{D,j}$.  We endow $X$ with the
symplectic form $\omega={i\over 2} \sum_i\left(da^i\wedge d\bar
a_{D,i}-da_{D,i}\wedge d\bar a^i\right)$ of type $(1,1)$ and also with
the holomorphic two-form $\omega_h =\sum_ida^i\wedge da_{D,i}$.  Then we
introduce arbitrary local coordinates $u^s,\,\,s=1,\dots, r$, on the
moduli space $\CM$, and describe a map $f:\CM\to X$ by functions
$a^i(u)$, $ a_{D,j}(u)$.  We require $f$ to be such that
$f^*(\omega_h)=0$; this precisely ensures that locally, if we pick
$u^i=a^i$, then $a_{D,j}$ must be of the form in \dualvar\ with some
holomorphic function $\CF$.  Then we take the metric on $\CM$ to be the
one whose Kahler form is $f^*(\omega)$; in formulas the metric is
\eqn\inform{
(ds)^2= \Im \sum_{s,t,i}{\partial a_{D,i}\over \partial u^s}
{\partial\bar a^{i} \over\partial \bar u^t} du^sd\bar u^t.}
If again we arrange $a,a_D$ as a $2r$-component column vector $v$,
then the formalism is invariant under transformations $v\to Mv+c$
with $M$ a matrix in $Sp(2r,{\bf R})$ and $c$ a constant vector.
Again, considerations involving  the charges will eventually require
that $M$ be in $Sp(2r,{\bf Z})$ and impose restrictions on $c$.

\subsec{Physical Interpretation Via Duality}

So far we have seen that the spin zero component of the $N=2$ multiplet
has a Kahler metric of a very special sort, constructed using a
distinguished set of coordinate systems.  This rigid structure is
related by $N=2$ supersymmetry to the natural linear structure of the
gauge field.  We have found that, for the spin zero component, the
distinguished parametrization is not completely unique; there is a
natural family of parametrizations related by $SL(2,{\bf R})$.  How does
this $SL(2,{\bf R})$ (which will actually be
reduced to $SL(2,{\bf Z})$) act on the gauge fields?

$SL(2,{\bf R})$ is generated by the transformations
\eqn\nameneeded{T_b = \pmatrix{ 1 & b \cr 0 & 1 \cr},~{\rm and}~ S=
\pmatrix{0 & 1\cr -1 & 0 } }
with real $b$.  The former acts as $a_D\to a_D+ba$, $a\to a$; this acts
trivially on the distinguished coordinate $a$, and can be taken to act
trivially on the gauge field.  By inspection of \ntweffln, the effect of
$a_D\to a_D+ba$ on the gauge kinetic energy is just to shift the
$\theta$ angle by $2\pi b$; in the abelian theory, this has no effect
until magnetic monopoles (or at least non-trivial $U(1)$ bundles) are
considered. Once that is done, the allowed shifts in the $\theta$ angle
are by integer multiples of $2\pi$; that is why $b$ must be integral
and gives essentially our first derivation of the reduction to
$SL(2,{\bf Z} )$.

The remaining challenge is to understand what $S$ means in terms of the
gauge fields.  We will see that it corresponds to electric-magnetic
duality.  To see this, let us see how duality works in Lagrangians of
the sort introduced above.

We work in Minkowski space and consider first
the purely bosonic terms involving only the gauge fields.
We use conventions such that
$F_{\mu\nu}^2= -({}^*F)_{\mu\nu}^2$ and ${}^{*}({}^*F) =-F$ where
${}^*F$ denotes the dual of $F$.  The relevant terms are
\eqn\someterms{{1\over 32 \pi} \Im \int \tau (a)\cdot (F + i {}^*F
)^2={1 \over 16\pi} \Im \int \tau (a)\cdot (F^2 + i {}^*FF ) \quad.}
Duality is carried out as follows.  The constraint $dF=0$ (which in the
original description follows from $F=dA$) is implemented by adding a
Lagrange multiplier vector field $V_D$.  Then $F$ is treated as an
independent field and integrated over.  The normalization is set as
follows.  The $U(1) \subset SU(2)$ is normalized such that all $SO(3)$
fields have integer charges (matter multiplets in the fundamental
representation of $SU(2)$ therefore have half integer charges).  Then,
a magnetic monopole corresponds to $\epsilon^{0\mu\nu\rho} \partial_\mu
F_{\nu\rho} = 8 \pi\delta^{(3)} (x)$.  For $V_D$ to couple to it with
charge one, we add to \someterms
\eqn\constterms{{1 \over 8\pi} \int V_{D\mu} \epsilon^{\mu\nu\rho\sigma}
\partial_\nu F_{\rho\sigma}  = {1 \over 8\pi} \int {}^*F_D F = {1\over
16 \pi}\Re \int ({}^*F_D  - iF_D)(F + i {}^*F) }
where $F_{D\mu \nu}=\partial_\mu V_{D\nu} - \partial_\nu V_{D\mu}$ is
the the field strength of $V_D$. We can now perform the Gaussian
functional integral over $F$ and find an equivalent Lagrangian for $V_D$,
\eqn\sometermsn{{1 \over 32 \pi} \Im {- 1 \over \tau } (F_D + i
{}^*F_D)^2 = {1\over 16\pi} \Im {- 1 \over \tau } (F_D^2 + i {}^*F_D
F_D) \quad .}

We now repeat these steps in $N=1$ superspace.  We treat $W_\alpha$ in
${1 \over 8\pi} \Im \int d^2 \theta \tau (A) W^2$ as an independent
chiral field.  The superspace version of the Bianchi identity $dF=0$ is
$\Im \CD W =0$ ($\CD$ is the supercovariant derivative).  It can be
implemented by a real vector superfield $V_D$ Lagrange multiplier.  We
add to the action
\eqn\lagconsts{{1 \over 4\pi} \Im \int d^4 x d^4 \theta V_D\CD W= {1
\over 4\pi} \Re \int d^4 x d^4 \theta i \CD V_D W= - {1 \over 4\pi} \Im
\int d^4x d^2 \theta W_DW   .}
Performing the Gaussian integral over $W$ we find an equivalent
Lagrangian
\eqn\gaugkinn{{1 \over 8 \pi} \Im \int d^2 \theta {- 1 \over \tau (A)}
W_D^2  .}

To proceed further, we need to transform the $N=1$ chiral multiplet $A$
to $A_D$.  The kinetic term
\eqn\chikin{\Im \int d^4 \theta h(A) \bar A }
is transformed by
\eqn\addef{A_D=h(A) }
to
\eqn\chikin{\Im \int d^4 \theta h_D(A_D) \bar A_D }
where $h_D(h(A))=- A $ is minus the inverse function.  Then using
$h^\prime(A)=\tau (A)$ the coefficient of the gauge kinetic term
\gaugkinn\ becomes
\eqn\gaugkinnd{-{ 1 \over \tau (A)}=- { 1 \over h^\prime (A)} =
h_D^\prime(A_D) = \tau _D(A_D) .}
Note that a shift of $h$ by a constant does not affect the Lagrangian.
Therefore, the duality transformation has a freedom to shift $A_D$ by a
constant.

The relations $A_D=h(A)$ and $h_D=- A$ mean that the duality
transformation precisely implements the missing $SL(2,{\bf Z})$
generator $S$.  The function $\tau =h^\prime$ is mapped by
\eqn\ffinal{\tau _D(A_D)= -{ 1 \over \tau (A)}}
Remembering that $\tau (a)={\theta (a) \over 2 \pi} + i {4 \pi \over
g(a)^2} $, we see that the duality transformation inverts $\tau $ rather
than the low energy gauge coupling $g(a)$.  (A similar phenomenon is
known in $R \leftrightarrow {1 \over R}$ duality in string theory
\ref\tduality{For a review, see A. Giveon, M. Porrati and E. Rabinovici,
RI-1-94, hep-th/9401139.},
where $B_{ij}+ i G_{ij} $, which is analogous to our $\tau$, is inverted
rather than $G_{ij}$.)

It is important to stress that unlike $\tau \rightarrow \tau +1$, the
duality transformation is not a symmetry of the theory.  It maps one
description of the theory to {\it another} description of the same
theory.

For other gauge groups $G$ the low energy Lagrangian has several abelian
fields, $A^i$, in the Cartan subalgebra.  Then
\eqn\sevgen{(A_D)_i= h_i (A^i)= \partial_i \CF(A^i) }
which leads to
\eqn\sevgenf{h_D^i(h_j(A^k))=  - A^i}
and the ``metrics''
\eqn\metdefi{\eqalign{
\tau _{ij}(A) = &\partial_i \partial_j \CF(A) = \partial_j h_i (A) \cr
\tau ^{ij}_D(A_D) = &\partial^i \partial^j \CF_D(A_D) = \partial^j
h_D ^i (A_D) \cr}}
satisfy
\eqn\metsat{f_{ij} f^{jk}_D= - \delta_i^k \quad .}
The above transformation together with the more obvious shifts
$A_{Di}\to A_{D_i}+M_{ij}A^j$ generate $Sp(2r,{\bf Z})$.

\subsec{Coupling To Gravity}

Before concluding this section, we would like to compare the structure
we have found to the ``special geometry'' that appears if the chiral
multiplet is coupled to $N=2$ supergravity \supgrat.  (This will not
be used in the present paper.)  In $N=2$
supergravity, the general Kahler metric for a system of $r$ chiral
superfields is described locally by a holomorphic function
$\CG_0(a^1,\dots,a^r)$ of $r$ complex variables $a^i$.  The Kahler
potential is
\eqn\gravpot{K_{\mit grav}=-\ln\left(2i(\CG_0-\bar\CG_0)+{i\over 2}
\sum_i\left(
\bar a^i{\partial \CG_0\over\partial a^i}-a^i{\partial\bar
\CG_0\over\partial \bar a^i}\right)\right).}
In global supersymmetry we had a local holomorphic function $\CF$
with
\eqn\flatpot{K={- i\over 2} \sum_i\left(
\bar a^i{\partial \CF\over\partial a^i}-a^i{\partial\bar
\CF\over\partial \bar a^i}\right).}
One would expect that there is some limit in which gravitational effects
are small and \gravpot\ would reduce to \flatpot.  How does this occur?

It suffices to set
\eqn\gravflat{\CG_0=-i{M_{\mit Pl}{}^2\over 4}+\CF,}
with $M_{\mit Pl}$ the Planck mass.  Then if $M_{\mit Pl}$ is much
larger than all relevant parameters, we get
\eqn\compgrav{K_{\mit grav}= -\ln M_{\mit Pl}{}^2+{ K\over M_{\mit
Pl}{}^2} +O(M_{\mit Pl}{}^{-4}). }
The constant term $-\ln M_{\mit Pl}{}^2$ does not contribute to the
Kahler metric, so up to a normalization factor of $1/M_{\mit Pl}^2$, the
Kahler metric with supergravity reduces to that of global $N=2$
supersymmetry as $M_{\mit Pl}\to \infty$ keeping everything else fixed.

More fundamentally, we would like to compare the allowed monodromy
groups.  In supergravity, the global structure is exhibited as follows.
One introduces an additional variable $a^0$ and sets $\CG=(a^0)^2\CG_0$.
One also introduces $a_{D,j}=\partial \CG/\partial a^j$ for
$j=0,\dots,r$.  Then one finds that the special Kahler structure of
\gravpot\ allows $Sp(2r+2,{\bf R})$ transformations acting
on $(a_{Di},a^j)$.\foot{Once once considers the gauge fields, this is
reduced to
$Sp(2r+2,{\bf Z})$.  The symplectic form preserved by $Sp(2r+2,{\bf R})$
is the usual one $\sum_ida^i\wedge
da_{D,i}$.}  Now, in decoupling gravity, we consider
$\CG$ to be of the special form in \gravflat.  In that case,
\eqn\adflat{a_{D,0}=-i{M_{\mit Pl}\over 2}.}
The other $a^i, a_{D,j}$ are independent of $M_{\mit Pl}$.  To preserve
this situation in which $M_{\mit Pl}$ appears only in $a_{D,0}$, we must
consider only those $Sp(2r+2,{\bf R})$ transformations in which the
transformations of all fields are independent of $a_{D,0}$.  These
transformations all leave $a^0$ invariant.  There is no essential loss
then in scaling the $a$'s so that $a^0=1$.  Arrange the $a_{D,i}, a^j$
with $i,j=1\dots r$ as a column vector $v$.
The $Sp(2r+2,{\bf R})$
transformations that leave invariant $a^0 = 1$ act on $v$
by $v\to Mv+c$ where $M\in Sp(2r,{\bf R})$ and $c$ is a
constant.  (The transformations with $c\not= 0$ do not leave $a_{D,0}$
invariant, but its variation is independent of $M_{\mit Pl}$ and so is
negligible in the limit in which gravity is weak.)  This is precisely
the duality group that we found in the global $N=2$ theory.

\newsec{Dyon masses}

The $SU(2)$ gauge theory under discussion has electrically and
magnetically charged particles whose masses satisfy
\eqn\clasmassd{M^2= 2 |Z|^2}
with
\eqn\zdefe{Z_{cl}= a(n_e + \tau_{cl} n_m) }
($\tau_{cl}={\theta\over 2\pi}+ i {4\pi \over g^2}$)
where $n_e$ and $n_m$ are the electric and magnetic charges; they are
integers as long as all elementary fields are in representations of
$SO(3)$ (fields that are in $SU(2)$ representations of half-integral
spin have half-integral $n_e$).  The origin of this formula \wo\ is that
$Z$ arises as a central extension in the $N=2$ supersymmetry algebra and
equation \clasmassd\ follows from the representation theory of $N=2$ for
``small'' representations.  One uses the algebra to show that
for any state of given $(n_m,n_e)$,
\eqn\bogbound{M\geq \sqrt 2|Z| }
with equality precisely for the ``small'' representations of $N=2$
(four helicity states instead of sixteen).  We will not review the
argument here.  States saturating the inequality are called
BPS-saturated states.

As stressed in \wo, the same interpretation should apply quantum
mechanically, but the coefficients of $n_e$ and $n_m$ in $Z$ might be
modified.  One way to find the modification is to calculate the central
extension of the algebra from the the low energy effective Lagrangian
\ntweffl.  This leaves an ambiguity in shifting $A$ and $A_D$ by a
constant.  As remarked above, such an ambiguity can be resolved by
considering the full high energy theory in its effective form as in
\ntweffln.

Alternatively, we can couple the theory based on \ntweffl\ to a
hypermultiplet -- two $N=1$ chiral multiplets $M$ and $\tilde M$ --
with electric charge $n_e $ and a canonical kinetic term.  As in
\ntwosup, the coupling to the gauge field is $N=2$ supersymmetric only
with a superpotential
\eqn\necoup{\sqrt{2} n_e A M \tilde M \quad.}
There is also a possibility of adding a mass term; this has the effect
of shifting $n_e A$ by a constant corresponding to the ambiguity
mentioned above.  By embedding the theory in a higher energy theory such
as \ntweffln,  this ambiguity is removed, and one learns in the pure gauge
theory that the coefficient of $M\tilde M$ in the superpotential
is precisely $\sqrt 2 n_e A$, with no additional additive constant.
With additional matter multiplets included
\nnext,  such a term will arise and play an important role.

For non-zero $a$, the fields $M$ and $\tilde M$ are massive.  As the
corresponding states are
in a ``small'' representation,
their mass is determined by the central extension in the algebra
\wo.  Comparing \clasmassd\ and \necoup, we conclude that $Z=a n_e$.
Using the duality transformation, it is clear that for magnetic
monopoles with magnetic charge $n_m$, $Z= a_D n_m$
and for dyons
\eqn\newzd{Z= a n_e + a_D n_m  .}
These formulas can be verified directly; for instance, to study the
BPS bound for monopoles, we consider the full high energy theory
in its effective form \ntweffln\ and
examine the bosonic terms in the Hamiltonian of a
magnetic monopole for real $\phi^a$:
\eqn\boson{\eqalign{
E= & {1 \over 4\pi} \Im  \int d^3 x\left(
 \tau_{ab} (D_i \phi)^a (D_i \phi)^b
+ \half \tau_{ab} B_i^aB_i^b\right) \cr
=&{1 \over 4 \pi} \Im \int d^3 x  \left(\left( { - 1 \over \tau}
\right)^{ab}
(D_i h)_a (D_i h)_b + \half \tau_{ab} B_i^aB_i^b \right)\cr
= &{1 \over 4 \pi} \Im  \left\{
\int d^3 x \left( { - 1 \over \tau} \right)^{ab}
\left[ (D_i h)_a \pm {1 \over \sqrt {2}} \tau_{ac} B_i^c \right] \left[
(D_i h)_b \pm {1 \over \sqrt {2}} \tau_{bd} B_i^d \right]\right\} \cr
& \qquad \mp \sqrt{2} \partial_i ( B_i^a h_a)  \cr
\ge & \left |{\sqrt{2} \over 4\pi}\oint d^2s B_i^a h_a \right| =
\sqrt{2} |n_m a_D|.  \cr}}
This inequality confirms that for monopoles $Z=n_ma_D$.

We would like to make a few comments:

\lfm{1.} The  expression \newzd\
has the correct semiclassical limit and manifest
duality.

\lfm{2.} Unlike \zdefe, it is renormalization group invariant.  Note
that it does not differ from \zdefe\ merely by replacing $i{ 4\pi \over
g^2}$ by $\tau (a)$; i.e.\ by the running coupling.  The distinction
between them appears already at one loop.

\lfm{3.} From this expression, it follows that the renormalization of
the BPS formula vanishes for $N=4$ supersymmetric Yang-Mills
theory.  For $N=4$, the classical expression \clasval\ is exact
(note the comment at the end of section 2) so there are no corrections
to the classical formula $a_D=\tau_{cl}a$.

\lfm{4.} The generalization of \newzd\ to an arbitrary gauge group is
\eqn\newzda{Z= a^i n_{e,i} +  h_i(a) n_m{}^i= a\cdot n_e + a_D \cdot
n_m }
where $a^i$ are local coordinates on the quantum moduli space and
$n_{e,i},\,n_m{}^i$ are the electric and magnetic charges.

Now let us discuss the restrictions that  the dyon mass formula puts on
the duality discussed in section 3.  First of all, it is
clear that $Z$, since it determines particle masses (or appears in the
supersymmetry algebra) must be invariant under the monodromies.  In
section 3, we arranged $(a_D,a)$ as a column vector $v$, and found that
the analysis of the Kahler metric permitted monodromies $v\to Mv+c$.
Since $n_ea+n_ma_D$ is not invariant under addition of a constant to $a$
or $a_D$, and there is no way to compensate for this by any
transformation of $n_e$ or $n_m$, we must set $c=0$. \foot{When matter
is included, additional terms appear in $Z$ and one no longer gets $c=0$
\nnext.} Moreover, under $v\to Mv$, we need $w\to wM^{-1}$ where $w$ is
the row vector $w=(n_m,n_e)$.  But since $(n_m,n_e)$ are integers,
$M^{-1}$ must be integer-valued.  As the determinant of $M$ is 1, it
follows that also $M$ is integer-valued; hence the monodromy group is at
most $SL(2,{\bf Z})$.

{\it Stability of BPS-Saturated States}

Now we will discuss the stability of BPS-saturated states.
Many of the following remarks are well known and none are new.

A BPS-saturated state of given $(n_m,n_e)$ determines
a vector $Z=n_ma_D+n_ea$ in the complex plane.
Its mass is $\sqrt{2}$ times the
length of that vector.  According to \bogbound, all other states with
the same $(n_m,n_e)$ are heavier.  Assuming that the ratio $a_D/a$
is not real, the complex numbers $a$ and $a_D$ generate a lattice in the
complex plane, and $Z$ is a point in that lattice.

Let us analyze a possible decay process of a BPS-saturated state
$S$ with $Z=an_e+ a_D n_m$ and mass $M=\sqrt{2}|Z|$ to states $S_i$
with $Z_i=n_{m,i}a_D+n_{e,i}a$
and masses $M_i \ge \sqrt{2}|Z_i|$.  Since the charges
$(n_m,n_e)$ must be conserved, $Z=\sum_i Z_i$.  It is clear from the
triangle inequality that
\eqn\decbou{|Z| \le \sum_i |Z_i|}
and hence
\eqn\decboum{M \le \sum_i M_i.}
Of course if $M<\sum_iM_i$ the decay is impossible.
Equality is achieved in \decbou\
when and only when all the states are BPS-saturated
and all the vectors $Z$ and $Z_i$ are aligned, that is
$t_i=Z_i/Z$ is real and positive and $\sum_it_i=1$.
Assuming that the ratio of $a_D/a$ is not real, this is possible
only if the charge vectors $(n_m,n_e)$ of the initial particles
and $(n_{m,i},n_{e,i})$ of the final particles are proportional.
This in turn is possible only if
$n_m$ and $n_e$ are not relatively prime; i.e.\ $(n_m,n_e)=(qm,qn)$ with
integers $q$, $m$ and $n$.

Conversely, states with $(n_m,n_e)=(qm,qn)$
are in fact only neutrally stable against
decay to $q$ states with charges $(m,n)$.

Now what happens to these stable particles (BPS
saturated with $(n_m,n_e)$ relatively prime) as we vary some of the
parameters that determine the vacuum?
Such small changes in the vacuum can be described by emission of zero momentum
particles
-- the neutral $u$ quanta.  Possible emission of such particles
does not affect the argument for stability of the stable BPS
saturated states so those states must persist as the parameters are varied.

So far we have assumed that $a_D/a$ is not real.
As explained in \vafa, the above  argument fails if, upon varying
the parameters, $a_D/a$ passes through the real axis.\foot{Their formulation
is slightly more general; they do not assume
that the allowed values of $Z$ are integer linear
combinations of two basic numbers $a$ and $a_D$.}
When that happens, the two dimensional lattice generated
by $a_D$ and $a$ collapses to a one dimensional configuration,
and it becomes much easier for the triangle inequality to collapse
to an equality.  For instance, if $a_D/a$ is real and irrational,
there are infinitely many points $n_{m,1}a_D+n_{e,1}a$ on the segment
between $0$ and $Z=n_ma_D+n_ea$; letting $Z_1$ be one such point,
and $Z_2=Z-Z_1$,  a BPS-saturated state of given $Z$ is only
neutrally stable against decaying to possible BPS-saturated
particles of given $Z_1$
and $Z_2$.

Moreover -- and this is the main point --
it was shown in \vafa\ that at least in two dimensions,
a BPS-saturated particle $S$ of given $(n_m,n_e)$
can disappear (or appear) when one passes through the point in parameter space
at which $a_D/a$ is real.   What happens is that, for, say,
${\rm Im}(a_D/a)>0$,
the $S$ particle, if it exists, is stable against decay to, say, $S_1+S_2$, but
for ${\rm Im}(a_D/a)\to 0$, the mass of the $S$ particle goes up to
the $S_1+S_2$ threshold.  Varying the parameters still further,
to ${\rm Im}(a_D/a)<0$, the $S$ particle no longer exists -- it has decayed
to $S_1+S_2$.  A BPS-saturated state of the given $(n_m,n_e)$ would
again be stable when ${\rm Im}(a_D/a)<0$, but such a particle may not exist.
In \vafa, the precise number of BPS-saturated states that appear
or disappear in this way was computed in two dimensions.  An analogous
computation in four dimensions would be desirable.

In the present paper, discontinuity of the BPS-saturated
spectrum will be found (in section 6) only for strong coupling where
it is difficult to explicitly check what is going on.  In a subsequent
paper \nnext, we will see such jumping also for weak coupling.

Such jumping does not occur for the $N=4$ theory since then
one has the exact formula $a_D=\tau_{cl}a$
ensuring that $a_D/a$ is not real.  Consequently, the spectrum of
BPS-saturated states of given $(n_m,n_e)$ is independent of
the coupling and so can be computed semiclassically for weak coupling.
This has, in fact, been assumed in tests of Olive-Montonen duality
for $N=4$.

\newsec{Structure Of The Moduli Space}

In section 2, we developed the general local framework for the low
energy effective action of the $N=2 $ theory.  At the outset of section
3, we noted that this framework could not be satisfactory globally
because the metric on the moduli space of vacua could not be positive
definite.  Instead, we found that the global structure could involve
certain monodromies; as we have explained,
the group generated by the monodromies is a subgroup of
$SL(2,{\bf Z})$.

\subsec{The Singularity at Infinity}

It is actually quite easy to see explicitly the appearance of
non-trivial monodromies.  In fact, asymptotic freedom implies a
non-trivial monodromy at infinity.  The renormalization group corrected
classical formula $\CF_{\mit{one \,\,loop}}=i\CA^2\ln(\CA^2
/\Lambda^2)/2\pi$ gives for large $a$
\eqn\asform{a_D = {\partial \CF\over\partial a}\approx {2ia\over
\pi}\ln(a/\Lambda) +{ia\over \pi}.}
It follows that $a_D$ is not a single-valued function of
$a$ for large $a$.  If we recall that the physical parameter is really
$u=\half a^2$ (at least for large $u$ and $a$), then the monodromy can be
determined as follows.  Under a circuit of the $u$ plane at large $u$,
one has $\ln u\to \ln u +2\pi i$, and hence $\ln a\to \ln a+\pi i$.  So
the transformation are
\eqn\monodromy{\eqalign{
&a_D\to - a_D+ 2a\cr
&a \to -a. }}
Thus, there is a non-trivial monodromy at infinity in the $u$ plane,
\eqn\ppmat{M_\infty= PT^{-2} = \left(\matrix{-1 & 2 \cr 0 & -1}\right) }
where $P$ is the element $-1$ of $SL(2,{\bf Z})$ and as usual
\eqn\asus{T=\pmatrix{ 1 & 1\cr 0 & 1\cr}.}

The factor of $P$ in the monodromy exists already at the classical
level.  As we said above, $a$ and $-a$ are related by a gauge
transformation (the Weyl subgroup of the $SU(2)$ gauge group) and
therefore we work on the $u$ plane rather than its double cover, the $a$
plane.  In the anomaly free ${\bf Z}_8$ subgroup of the $R$ symmetry
group $U(1)_\CR$, there is an operation that acts on $a$ by $a\to -a$;
when combined with a Weyl transformation, this is the unbroken symmetry
that we call $P$.     Up to a gauge transformation it acts on the bosons
by $\phi\to -\phi$, so it reverses
the sign of the low energy electromagnetic field which in terms of
$SU(2)$ variables is proportional to $\Tr\, (\phi F)$.
Hence it reverses the signs of all electric and magnetic charges and
acts as $-1\in SL(2,{\bf Z})$.
The $P$ monodromy could be removed by
(perhaps artificially) working on the $a$ plane instead of the $u$
plane.

The main new point here is the factor of $T^{-2}$ which arises at the
quantum level.  This factor of $T^{-2}$ has a simple physical
explanation in terms of the electric charge of a magnetic monopole.
As explained in
\ref\moncharge{E. Witten, \pl{86}{1979}{283}.},
magnetic monopoles labeled by $(n_m,n_e)$ have anomalous electric
charge $ n_e+{\theta_{{\mit eff}}\over 2 \pi} n_m$.  The appropriate
effective theta parameter is the low energy one
\eqn\lowener{\theta_{{\mit eff}}=2 \pi \,\Re \tau(a)= 2 \pi \,\Re { d a_D
\over d a} =  2 \pi \,\Re { d a_D/d u \over d a
/d u}.}
For large $|a|$, we have $\theta_{{\mit eff}} \approx -
4{\rm arg}(a)$ which
can be understood from the anomaly in the $U(1)_{\CR}$ symmetry.
The monodromy at infinity transforms the row vector
$(n_m,n_e) $ to $(-n_m,-n_e-2n_m)$, which implies that $(a_D,a)$
transforms to $(-a_D+2a,-a)$.  The electric
charge of the magnetic monopole can in fact be seen in the formula for
$Z$, which if we take $a_D$ from \asform\ and set $a=a_0 e^{-
i\theta_{\mit eff}/4}$ (with $a_0>0$) is
\eqn\zeff{Z \approx a_0 e^{-
i\theta_{\mit eff}/4}\left\{ \left(n_e+{\theta_{\mit eff} n_m\over
2\pi}\right)+ in_m\left( {2\ln a_0/\Lambda + 1\over \pi}\right)
\right\}.}
The monodromy under $\theta_{\mit eff} \to \theta_{\mit eff} + 4\pi$ is
easily seen from this formula to transform $(n_m,n_e)$ in the expected
fashion.  Of course, this simple formula depended on the semiclassical
expression \asform\ for $a_D$; with the exact expressions we will
presently propose, the results are much more complicated, in part
because the effective theta angle is no longer simply the argument of
$a$.

\subsec{Singularities at Strong Coupling}

The monodromy at infinity means that there must be an additional
singularity (or topological complication) somewhere in the $u$ plane.
If ${\cal M}'$ is the moduli space of vacua with all singularities
deleted, then the monodromies must give a representation of the
fundamental group of ${\cal M}'$ in $SL(2,{\bf Z})$.  Can this
representation be abelian?  If the monodromies all commute with $PT^{-2}$,
then $a^2$ is a good global complex coordinate, and the metric is
globally of the form \locform\ with a global harmonic function $\Im
\tau(a)$.  As we have already noted, such a metric could not be
positive.

The alternative is to assume a nonabelian representation of the
fundamental group.  This requires at least two more punctures of the $u$
plane (in addition to infinity).  Since there is a symmetry $u
\leftrightarrow -u$ acting on the $u$ plane, the minimal assumption is
that there are precisely two more punctures exchanged by this symmetry.
In this paper, we will find that this assumption leads to a unique and
elegant solution that passes many tests.  (In a following paper \nnext,
we will in some sense derive this assumption from more general
properties of $N=2$ systems with matter.)

The most natural physical interpretation of singularities in the $u$
plane is that some additional massless particles are appearing at a
particular value of $u$.  Such a phenomenon of singularities in
moduli space associated with the occurrence of extra
massless particles has already been observed in
$N=1$ theories \moduli\ and we will argue that it also
happens in our problem.

For instance, in the classical theory, at $u=a=0$, the $SU(2)$ gauge
symmetry is restored; all the gluons become massless.  In fact
classically $a_D=4\pi  ia/g^2$ also vanishes at this point, and the
monopoles and dyons become massless as well.  One might be tempted to
believe that the missing singularity comes from an analogous point in
the quantum theory at which the gauge boson masses vanish.  Though this
behavior might seem unusual in asymptotically free theories in general,
there are good indications that some $N=1$ theories have an infrared
fixed point with massless nonabelian gluons
\nref\newseiberg{N. Seiberg, to appear.}%
\refs{\moduli,\newseiberg}.

However, there are good reasons to doubt that the $N=2$ theory has this
behavior.  First of all, to make sense of the monodromies, one needs (as
we saw above) not a single singularity but (at least) a pair of
singularities at non-zero $\langle \Tr\,\phi^2\rangle$.  One might be
willing to believe that $\langle \Tr\,\phi^2\rangle \not= 0$ in the
theory with massless nonabelian gauge bosons because of spontaneous
breaking of the discrete chiral symmetry.  This assumption, however,
clashes with the asymptotic conformal invariance that one would expect
in the infrared if the gauge bosons are massless.  In fact, a non-zero
expectation value of $\Tr\,\phi^2$ contradicts conformal invariance
unless $\Tr\,\phi^2$ is of dimension zero.  In a unitary quantum field
theory, the only operator of zero dimension is the identity;
$\Tr\,\phi^2$ cannot mix with the identity under renormalization because
it is odd under a global symmetry.

Moreover, a conformally invariant point for the $N=2$ theory is
far-fetched because conformal invariance together with $N=2$
supersymmetry implies invariance under the full $N=2$ superconformal
algebra including the $U(1)_\CR$ symmetry.  Thus, the instanton anomaly
in the $U(1)_\CR$ symmetry would have to somehow disappear.  Moreover,
for a field such as $\CO=\Tr\,\phi^2$ which is in a chiral multiplet,
superconformal invariance implies that the dimension $D(\CO)$ and
$U(1)_\CR$ charge $\CR(\CO)$ are related by $D(\CO)=\CR(\CO)/2$.  Thus
at a conformal point the dimension of $\Tr\,\phi^2$ would have the
canonical value $2$, not the value 0 that it should have in order to
have an expectation value.

\subsec{Interpretation Of The Singularities}

Since the above discussion of massless gauge bosons does not appear
promising, we will assume that the singularities come from massive
particles of spin $\leq 1/2$ that become massless at particular points
in the moduli space.  Since there are no such elementary multiplets,
these must be bound states or collective excitations.
It might sound counter intuitive that such objects can
become massless.  However, a similar phenomenon of massless bound states
(which are not Goldstone bosons) has been observed in $N=1$ theories
\moduli\ and we will argue that it also happens in $N=2$.

The possibilities are severely restricted by the structure of $N=2$
supersymmetry: a massive multiplet of particles of spins $\leq 1/2$ must
be a hypermultiplet that saturates the BPS bound.

In the semiclassical approximation the only such hypermultiplets in the
$N=2$ gauge theory are the monopoles and dyons whose mass
renormalization was the subject of section 4.  We will interpret the
needed singularities as arising when these particles become massless.
For instance, from the discussion of masses in section 4, the monopole
becomes massless, while the gluons remain massive, at a point where
$a_D=0$ while $a\not= 0$.  Similarly a $(1,1)$ dyon becomes massless if
$a+a_D=0$ while $a,a_D\not= 0$.

Proposing that magnetic monopoles become massless and dominate the low
energy landscape (at certain points in the moduli space of vacua)
may seem bold.  In the rest of this paper we will give evidence for
this hypothesis as follows:

\lfm{1.} We will use the renormalization group
to compute the $SL(2,{\bf Z})$ monodromy that arises near a point
at which a hypermultiplet becomes massless.
We will then show that if the hypermultiplets that are relevant
are the monopoles and dyons that are visible semiclassically, then
the monodromies work out consistently.

\lfm{2.} The underlying $N=2$ theory can be perturbed to an $N=1$
theory by adding $\Tr\,\Phi^2$ as a superpotential.  It is believed that
this causes confinement of quarks.  We will see that the same
perturbation, added near the point at which monopoles are becoming
massless, causes the monopoles to condense (develop a vacuum expectation
value).  This naturally leads to confinement of charges, giving -- for
the first time -- an example in which confinement in a nonabelian theory
is naturally understood in terms of monopole condensation.

\lfm{3.} Finally, we will show that if it is assumed that there is a minimal
set of singularities coming from massless monopoles and dyons, then one
can determine uniquely and exactly the full structure of the low energy
theory, including the Kahler metric on the moduli space of vacua and the
particle masses as a function of the parameters.  In particular the
puzzle with positivity of the metric mentioned at the outset of section
3 is naturally overcome.

\vskip.5cm
Explaining these points will occupy the rest of this paper.

\subsec{Effects Of A Massless Monopole}

Our first task is to analyze the behavior of the effective Lagrangian
near a point $u_0$ on the moduli
space where magnetic monopoles become massless, that is, where
\eqn\adofuz{a_D(u_0)=0.}
Since monopoles couple in a non-local way to the original photon, we
cannot use that photon
in our effective Lagrangian.  Instead, we should perform a
duality transformation and write the effective Lagrangian in terms of
the dual vector multiplet $\CA_D$.  The low energy theory is therefore
an abelian gauge theory with matter (an $N=2$ version of QED).  The
unusual fact that the light matter fields are magnetically charged
rather than electrically charged does not make any difference to the low
energy physics.  The only reason we call these particles monopoles
rather than electrons is that this language is appropriate
at large $|u|$ where the theory is
semiclassical.

The dominant effect on the low energy gauge coupling constant is due to
loops of light fields.  In our case, these are the light monopoles.
The low energy theory is not asymptotically free and therefore its gauge
coupling constant becomes smaller as the mass of the monopoles becomes
smaller.  Since the mass is proportional to $a_D$, the low energy
coupling goes to zero as $u \rightarrow u_0$.  The electric coupling
constant which is the inverse of the magnetic one diverges at that
point.

More quantitatively, using the one loop beta function, near the point
where $a_D=0$, the magnetic coupling is
\eqn\fdnearad{\tau_D \approx - {i \over \pi} \ln a_D .}
Since $a_D$ is a good coordinate near that point,
\eqn\apprxadu{ a_D\approx c_0(u-u_0)}
with some constant $c_0$.  Using $\tau_D=d h_D /d
a_D$, we learn that
\eqn\anearad{a(u) = - h_D (u) \approx a_0 + {i \over  \pi} a_D \ln a_D
\approx a_0 + {i \over \pi } c_0 (u-u_0) \ln (u - u_0) }
for some constant $a_0=a(u=u_0)$. This constant $a_0$ cannot be zero
because if it had been zero, all the electrically charged particles
would have been massless at $u=u_0$ and the computation using
light monopoles only would not be valid.  (In fact, there would
be no local effective field theory incorporating both the light
electrons and light monopoles.)

Now we can read off the monodromy.  When $u$ circles around
$u_0$, so $\ln(u-u_0)\to \ln(u-u_0)+2\pi i$, one has
\eqn\monf{\eqalign{  a_D & \to a_D\cr
                     a & \to a-2a_D .\cr}}
This effect is a sort of dual of the monodromy at infinity.
Near infinity, the monopole gains electric charge, and near $u=u_0$,
the electron gains magnetic charge.  (It does not come back as a dyon
but as a pair of particles for reasons explained at the end of section
6.)  \monf\ can be represented by the $2\times 2$ monodromy matrix
\eqn\monmat{M_1=ST^2S^{-1}= \pmatrix{ 1 & 0 \cr -2 & 1}.}

\subsec{The Third Singularity}

With our assumption that there are only three singularities (counting
$u=\infty$) and with two of the three monodromies determined in \ppmat\
and \monmat, we can now determine the third monodromy, which we will
call $M_{-1}$.  (The motivation for the notation is that we will
eventually introduce parameters in which the singularities at finite $u$
are at $u=1$ and $u=-1$.)  With all of the monodromies taken in the
counter clockwise direction as in figure 1, the monodromies must obey
$M_1M_{-1}=M_\infty$, and from this we get
\eqn\lastmon{M_{-1}= (TS)T^2(TS)^{-1}= \pmatrix{ -1 & 2 \cr -2 & 3}.}

\fig{The $u$ plane with monodromies around $1$, $-1$, and $\infty$.
Note the choice of base point in the definition of the monodromies.}
{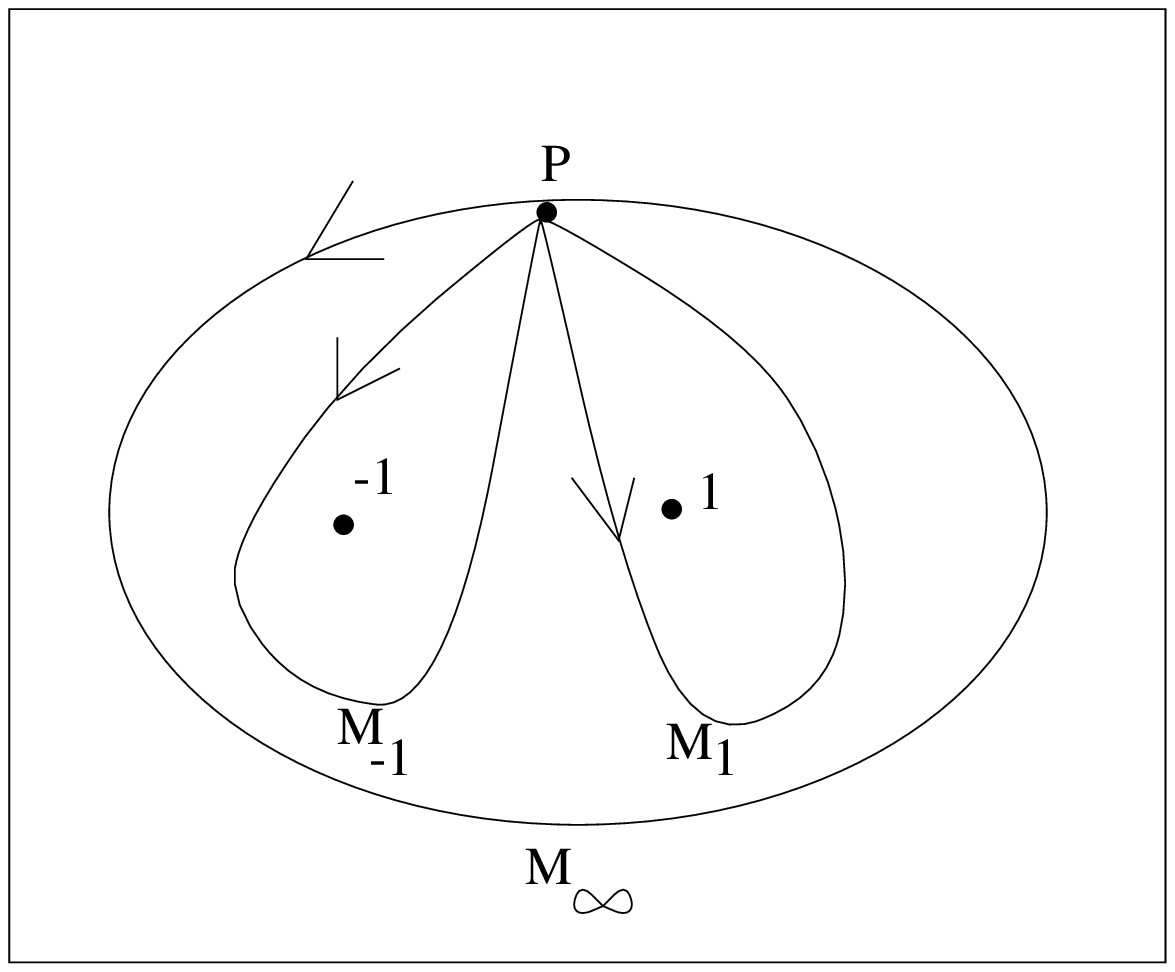}{10 truecm}
\figlabel\igure
\bigskip

The matrix $M_{-1}$ is conjugate to $M_1$.  In fact, if
\eqn\ifa{A=TM_1= \pmatrix{-1  &  1\cr -2 & 1},}
then
\eqn\conjma{M_{-1}= AM_1A^{-1}.}
Hence, $M_{-1}$ can arise from a massless particle, just like $M_1$.
\conjma\ would also hold
if $A$ is replaced by $AM_1{}^r$ for any integer $r$.

What kind of particle should become massless to generate this
singularity?  If one arranges the charges as a row
vector $q=(n_m,n_e)$, then the massless particle that produces a
monodromy $M$ has $qM=q$.  For instance, monodromy $M_1$ arises from
a massless monopole of charge vector $q_1=(1,0)$, and using the
known form of $M_1$, one has $q_1M_1=q_1$.  Duality symmetry implies
that this must be so not just for the particular monodromy $M_1$ but for
any monodromy coming from a massless particle.  Upon setting
$q_{-1}=(1,-1)$, we get $q_{-1}M_{-1}=q_{-1}$, and hence the monodromy
$M_{-1}$ arises from vanishing mass of a dyon of charges $(1,-1)$.

It seems that we are seeing massless particles of charges $(1,0)$
or $(1,-1)$.  However, there is in fact a complete democracy among
dyons.  The BPS-saturated dyons that exist semiclassically
have charges $(1,n)$ (or $(-1,-n)$) for arbitrary integer $n$.
The monodromy at infinity brings about a shift $(1,n)\to
(1,n-2)$.  If one carries out this shift $n$ times before proceeding
to the singularity at $u=1$ or $u=-1$, the massless particles producing
those singularities would have charges $(1,-2n)$ and $(1,-1-2n)$,
respectively.  This amounts to conjugating the representation of the
fundamental group by $M_\infty{}^n$.

The particular matrix $A$ in \ifa\
obeys $A^2=-1$, which is equivalent to the identity as an automorphism
of $SL(2,{\bf Z})$.  Conjugation by $A$ implements
the underlying ${\bf Z}_2$ symmetry of the quantum moduli space which
according to our assumptions exchanges
the two singularities.   The
${\bf Z}_2$ maps $M_1 \rightarrow M_1'=M_{-1}$, $M_{-1} \rightarrow
M_{-1}'=M_1$ and $M_\infty \rightarrow M_\infty'=M_1'M_{-1}'=
M_{-1}M_1$.  Note that $M'_\infty$ is not just obtained from
$M_\infty$ by conjugation, but the relation $M_\infty
=M_1M_{-1}$ is preserved.  The reason for that is that
(as in any situation in which one is considering a representation of
the fundamental group of a manifold in a nonabelian group), the definition
of the monodromies requires a choice of base point, as shown in
figure \igure.  The operation $u\to -u$ acts on the base point, and this
has to be taken into account in determining how $M_\infty$ transforms
under ${\bf Z}_2$.

One can go farther and show that if one assumes the existence of
a ${\bf Z}_2$ symmetry between $M_1$ and $M_{-1}$, then they
must be conjugate to $T^2$, and not some other power of $T$.
In our derivation of the monodromy \monmat, the 2 came from something
entirely independent of the assumption of a ${\bf Z}_2$ symmetry,
namely, from the charges and multiplicities of the monopoles
that exist semiclassically.

\subsec{Monopole Condensation And Confinement}

We will now explain a quite satisfying physical phenomenon
which was in fact at the heart of how
some of these things were originally discovered.

We recall that the underlying $N=2$ chiral multiplet $\CA$ decomposes
under $N=1$ supersymmetry as a vector multiplet $W_\alpha$ and a chiral
multiplet $\Phi$.  Breaking $N=2$ down to $N=1$, one can add a
superpotential $W= m\Tr\, \Phi^2$ for the chiral multiplet.  This gives a
bare mass to $\Phi$, reducing the theory at low energies to a pure $N=1$
gauge theory.  The low energy theory has a ${\bf Z}_4$ chiral symmetry.
This theory is strongly believed to generate a mass gap, with
confinement of charge and spontaneous breaking of ${\bf Z}_4 $ to ${\bf
Z}_2$.  Furthermore, there is no vacuum degeneracy except what is
produced by this symmetry breaking, so that there are precisely two
vacuum states
\ref\wittenin{E. Witten, \np{202}{1982}{253}.}.

How can this be mimicked in the low energy effective $N=2$ theory?  That
theory has a moduli space ${\cal M}$ of quantum vacua.  The massless
spectrum at least semiclassically consists solely of the abelian chiral
multiplet $\CA$ of the unbroken $U(1)$ subgroup of $SU(2)$.  If those
are indeed the only massless particles, the effect in the low energy
theory of turning on $m$ can be analyzed as follows.  The operator
$\Tr\, \Phi^2$ is represented in the low energy theory by a chiral
superfield $U$.  Its first component is the scalar field $u$ whose
expectation value is
\eqn\regar{\langle u \rangle = \langle \Tr\,\phi^2\rangle}
($\phi$ is the $\theta=0$ component of the superfield $\Phi$).  This
is a holomorphic function on the moduli space.
At least for small $m$ we should add to our low energy
Lagrangian an effective superpotential $W_{\mit eff}=m U$ (soon we will
show that this is the exact expression also for large $m$).

Turning on the superpotential $m U$ would perhaps eliminate almost
all of the vacua and in the surviving vacua give a mass
to the scalar components of $\CA$.  But if there are no extra degrees
of freedom in the discussion, the gauge field in $\CA$ would
remain massless.  To get a mass for the gauge field, as is needed since
the microscopic theory has a mass gap for $m\not= 0$, one needs either
(i) extra light gauge fields, giving a non-abelian gauge theory and
possible strong coupling effects, or (ii) light charged fields,
making possible a Higgs mechanism.

Thus we learn, as we did in discussing the monodromies, that somewhere
on $\CM$ extra massless states must appear.  The option (i) does not
seem attractive, for reasons that we have already discussed.  Instead we
will consider option (ii), with the further proviso, from our earlier
discussion, that the light charged fields in question are monopoles and
dyons.

Near the point at which there are massless monopoles, the monopoles
can be represented in an $N=1$ language
by ordinary (local) chiral superfields
$M$ and $\widetilde M$, as long as we describe the gauge field
by the dual to the original photon, $\CA_D$.  The superpotential is
\eqn\omigob{\widehat W= \sqrt 2 A_DM\widetilde M + m U(A_D), }
where the first term is required by $N=2$ invariance of the $m=0$ theory,
and the second term is the effective contribution to the superpotential
induced by the microscopic perturbation $m\,\Tr\,\Phi^2$.

The fact that the superpotential \omigob\ is exact can be established by
using the non-renormalization theorem of \nonren\ as follows.  For $m=0$
the theory is invariant under $SU(2)_R$.  It will suffice to consider
its $U(1)_J$ subgroup \vecsym.  This is an $N=1$ $R$ symmetry under which
$\Phi$ has charge zero.  The two $N=1$ chiral fields $M$ and $\tilde M$
are in an $N=2$ hypermultiplet.  Therefore, according to \chisym, they
both have charge one.  The presence of
a term $m \Tr \,\Phi^2$ in the microscopic
superpotential shows that the parameter $m$ carries charge two.  The low
energy superpotential is holomorphic in its variables $ \widehat W(m, M
\tilde M, A_D)$ and should have charge two under $U(1)_J$.  Imposing that
it is regular at $m=M\tilde M =0$, we find that it is of the form
$\widehat W= mf_1(A_D)+ M \tilde M f_2( A_D)$.  The functions $f_1$ and
$f_2$ are independent of $m$ and can be determined by examining the
limit of small $m$, leading to \omigob.

The low energy vacuum structure is easy to analyze.  Vacuum states
correspond to solutions, up to gauge transformation, of
\eqn\ubbo{d\widehat W=0}
that obey the additional condition
\eqn\dcond{|M|=|\widetilde M|}
(we denote by $M$ and $\tilde M$ both the
superfields and their first components).
The latter condition comes from vanishing of the $D$ terms.
Implementing these conditions, one finds if $m=0$ that vacuum states
correspond to $M=\tilde M=0$ with arbitrary $a_D$; this is simply the
familiar moduli space $\CM$.  If $m\not=0$ the result is quite
different.  We get
\eqn\vacstr{\eqalign{ \sqrt 2 M \tilde M + m{du\over da_D} & = 0 \cr
                           a_DM=a_D\tilde M & = 0 . \cr}}
Assuming that $du\not= 0$, the first equation requires $M,\tilde M \not
= 0$, whence the second equation requires $a_D=0$.  Imposing also
\dcond, we get a unique solution up to gauge transformation, with
\eqn\vacval{M=\widetilde M=\left(-mu'(0)/\sqrt 2\right)^{1/2}.}

Expanding around this vacuum, it is easy to see that there is a mass
gap.  For instance, the gauge field gets a mass by the Higgs mechanism,
since $M,\tilde M\not= 0$.  The Higgs mechanism in question is a
magnetic Higgs mechanism, since the fields with expectation values are
monopoles!  Condensation of monopoles will induce confinement of
electric charge.  Thus, we get an explanation in terms of the low energy
effective action of why the microscopic theory becomes confining when
the $m\,\Tr\,\Phi^2$ superpotential is added.

We have also noted that in the presence of the perturbing
superpotential, the microscopic theory has a ${\bf Z}_4$ global
symmetry, spontaneously broken down to ${\bf Z}_2$.  This symmetry
breaking is manifest in the effective theory, since the broken symmetry
exchanges the point where $a_D=0$ and there is a massless monopole with
a point where $a-a_D=0$ and there is a massless dyon.  Thus, the
effective theory has two vacuum states from the two points of extra
massless particles (corresponding to the two singularities in the $u$
plane that were discussed above), related by a broken symmetry, in
parallel with what is expected microscopically.

In the microscopic description, one attributes the properties of the two
vacuum states (for $m\not= 0$) to a difficult-to-understand strong gauge
coupling.  In the low energy theory, we have found a perfectly peaceful
description involving a {\it weakly coupled} theory of monopoles and
photons; the coupling constant flows to zero in the infrared if $m=0$,
and in general flows to a value of order $-1/\ln m$, since this is
the behavior of weakly coupled QED.  The original,
electric, gauge coupling is the inverse of the magnetic coupling, so
flows in the infrared to a value of order $- \ln m$; in this sense the
two vacua that survive when $m$ is small and non-zero are strongly coupled.

The effective superpotential \omigob\ gives a good description of the
low energy physics near the point $u_0$ where the monopoles are light.
What is its meaning far from $u_0$?
Then, except for the expectation values of the
monopoles and their masses, few of their properties
could be correctly inferred from the effective Lagrangian.
The effective Lagrangian has even greater difficulties if one tries
to continue it to $u=-u_0$ where massless dyons should appear.
There cannot be an effective field theory containing both the monopoles
and dyons as elementary fields, as they are not relatively local.
The theory we are discussing is an interesting example of a theory
in which, while in the neighborhood of any one vacuum there is a
good description by a low energy effective theory, there is no low
energy effective theory that is reasonable everywhere.

\subsec{The Effective Parameter}

Finally, we will tie up some remaining loose ends and at the same
time prepare for the solution of the model in the next section.

In the above, we found two vacuum states for $m\not= 0$,
while assuming that
$du \not= 0$.  We would get additional states, with $M=\tilde M=0$,
for any point at which $du =0$.  We do not want any
additional vacuum states, since two is the correct number of vacua
in the microscopic theory, so we assume that $du\not=0$
everywhere.

The fact that $du\not=0$ means that
$u$ is everywhere a good local coordinate on the space of vacua.  Is it
also a good coordinate globally?  The question is whether there
is precisely one vacuum for given $u$.  According to \regar,
$u$ is simply the expectation value of $\Tr\,\phi^2$, regarded
as a function on the space of vacua.  At least when $u= \Tr\phi^2$
is large, perturbation theory is reliable and there is precisely one
vacuum for given $u$.  This combination of facts strongly
suggests that $u$ is a good global coordinate on the space $\CM$ of
vacua.

In proceeding in the next section to propose a solution of the model,
we will assume that $\CM$ is just the $u$ plane (with $u$ equal to
the expectation value of $\Tr\,\phi^2$) with precisely the two
singularities that we proposed above.
We will pick a renormalization convention such that the singularities
in the $u$ plane are at 1 and $-1$.
Using the solution of the model that we will propose in the next section,
it would be possible to compare to perturbation theory and determine
how this convention compares to other conventions such as $\bar{MS}$.

\newsec{The Solution Of The Model}

In this section, we will, finally, put the pieces together,
with a couple of new ingredients, and make our proposal
for the solution of the model.

The moduli space $\CM$ of quantum vacua is to be the $u$ plane
with singularities at $1,-1$, and $\infty$ and a ${\bf Z}_2$
symmetry acting by $u\leftrightarrow -u$.  Duality means that over
the punctured $u$ plane there is a flat $SL(2,{\bf Z})$ bundle $V$
with the following monodromies around $\infty, 1, $ and $-1$:
\eqn\folmon{\eqalign{M_\infty & = \pmatrix{-1 & 2 \cr
                                            0 & -1\cr}\cr
                     M_1      & = \pmatrix{1 & 0 \cr
                                           - 2 & 1\cr} \cr
                     M_{-1}  & = \pmatrix{-1 & 2 \cr
                                          -2 & 3 }.\cr}}
The quantities $(a_D(u),a(u))$ are a holomorphic section
of the bundle $V$ (or of $V\otimes {\bf C}$ to be precise).
This section is to be determined by its asymptotic behavior.
Near $u=\infty$,
\eqn\uinfty{\eqalign{ a\approx & \sqrt {2u} \cr
                      a_D\approx & i {\sqrt{2u} \over \pi } \ln u .\cr}}
Near $u=1$,
\eqn\uone{\eqalign{a_D \approx & c_0(u-1)\cr
                   a   \approx & a_0 + {i \over \pi}a_D\ln a_D,\cr}}
with constants $a_0,c_0$.
The behavior near $u=-1$ is similar, with $a-a_D$ replacing $a_D$.

There is also one more important constraint.  The metric
on the $u$ plane can be written
\eqn\dss{(ds)^2 ={\rm Im}(\tau)\cdot |da|^2}
with
\eqn\deftau{\tau={da_D/du\over da/du.} }
To ensure positivity of the metric, ${\rm Im}(\tau)$
must be positive definite.  We need not specify the asymptotic
behavior of $\tau$ near $-1,1$, and $\infty$, since
this is determined by the asymptotic behavior of $a$ and $a_D$.

The key point in making it practical to solve these conditions
is that the flat $SL(2,{\bf Z})$ bundle has a nice interpretation.
To begin with, note that the monodromy matrices in equation \folmon\ are all
congruent to $1$ modulo 2.  So these matrices do not generate
$SL(2,{\bf Z})$ -- at most they could generate the subgroup of
$SL(2,{\bf Z})$ consisting of matrices congruent to 1 modulo 2.
This subgroup is usually called $\Gamma(2)$.
In fact, $M_\infty$ and $M_{1}$ do generate $\Gamma(2)$.\foot{For
a quick proof of this and other assertions
made presently, see pp. 92-3 of
\ref\clemens{C. H. Clemens, {\it A Scrapbook Of Complex Curve Theory}
(Plenum Press, 1980).}.}
Moreover, the $u$-plane punctured at $1,$ $-1$, and $\infty$
has a very special interpretation.  It can be thought of
as the quotient of the upper half plane $H$ by $\Gamma(2)$.
Indeed, as $\Gamma(2)$ is of index six in $SL(2,{\bf Z})$,
the quotient $H/\Gamma(2)$ is a six-fold cover of the usual
modular domain.  This quotient has three cusps, which we can take to be
at $1,-1$, and $\infty$, with precisely the monodromies
in \folmon.

The family of curves parametrized by $H/\Gamma(2)$ can be described very
explicitly, by the equation
\eqn\tcover{y^2 = (x-1)(x+1)(x-u).}
The idea here is that for every $u$, there is a genus one Riemann
surface $E_u$, determined by equation \tcover.  This equation describes
a double cover of the $x$ plane branched over $-1,1,\infty$, and $u$.
The curve $E_u$ becomes singular when (and only when) two branch points
coincide.  This occurs precisely for $u=1,-1$, or $\infty$.

Note among other things that \tcover\ has a manifest symmetry $w$
that maps
$u\to -u$, $x\to -x$, $y\to \pm iy$.
This generates a ${\bf Z}_4$ symmetry, but only a ${\bf Z}_2$ quotient
acts on the $u$ plane.  Indeed, $P=w^2$ acts trivially on $u$ and $x$ while
mapping $y\to -y$.  It will turn out that $a$ and $a_D$ are odd under
$y\to -y$ and so odd under $P$.  The symmetry structure just described
is precisely that of the field theory that we are aiming to solve.

The Riemann surface $E_u$ has a two dimensional first homology group
$V_u=H^1(E_u,{\bf C})$.  The $V_u$ are fibers of a flat bundle $V$ over
the punctured $u$ plane, and this is the bundle of which the
pair $(a_D,a)$ is a section.
This is a convenient description of the bundle, as we will see.
The bundle $V$ can be trivialized
locally by picking a pair of independent and continuously varying
one-cycles $\gamma_1,\gamma_2$
on $E_u$; these can be normalized so that the intersection number is
\eqn\innumb{\gamma_1\cdot \gamma_2=1.}

The space $H^1(E_u,{\bf C})$ can be thought of as the space of
meromorphic $(1,0)$-forms on $E_u$ of vanishing residues,
modulo exact forms (or total derivatives).  The heuristic idea
is that if $\lambda$ is such a one-form, then it represents an element
of $H^1(E_u,{\bf C})$ because it can be paired with one-cycles by
\eqn\paircy{\gamma\to \oint_\gamma \lambda.}
The condition that the residues of $\lambda$ vanish ensures
that this pairing is invariant under deformation
of $\gamma$ even across a pole of $\lambda$.  One identifies
$\lambda\sim\lambda +dw$ (with $w$ a meromorphic function) since
the exact differential $dw$ would not contribute to the contour integral
\paircy.

As $H^1(E_u,{\bf C})$ is two-dimensional, a basis is provided by any
two linearly independent elements, for instance
\eqn\forinst{\lambda_1={dx\over y} ~{\rm and}~\lambda_2={x\,\,dx\over
y}.}
Here $\lambda_1$ is actually a holomorphic differential, having no poles
even at infinity; it is up to a scalar multiple the unique
holomorphic differential on $E_u$.  (In terms of the Hodge decomposition
of $H^1(E_u,{\bf C})$, it represents an element of $H^{1,0}$.)
If one picks on $E_u$ a basis of one-cycles normalized as in \innumb,
then the periods
\eqn\theper{b_i=\oint_{\gamma_i}\lambda_1,\,\,\,i=1,2}
obey
\eqn\theyob{{b_1\over b_2}=\tau_u,}
with $\tau_u$ the usual $\tau$ parameter of the elliptic curve $E_u$,
which has the fundamental property
\eqn\imtau{{\rm Im}(\tau_u) >0.}
Under a change in the basis of $\gamma$'s, $\tau_u$ would be transformed
by the standard action of $SL(2,{\bf Z})$ on the upper half plane.  As
for $\lambda_2$, its only pole is a double pole at infinity (where the
residue vanishes, since in any case the sum of the residues of a
meromorphic differential is zero).

As \forinst\ gives a basis of $H^1(E_u,{\bf C})$, an arbitrary section
of the flat bundle $V_u$ can be represented in the form
\eqn\repform{\lambda=a_1(u)\lambda_1+a_2(u)\lambda_2.}
This representation is not necessarily convenient, as it may be
convenient to use the freedom of adding $dw$ for some function $w$.  In
any event, once a section $\lambda$ is chosen, to extract the components
$a_D$ and $a$ of this section, one picks the basis $\gamma_1,\gamma_2$
of one-cycles and defines
\eqn\thendef{\eqalign{a_D & =\oint_{\gamma_1}\lambda\cr
                      a &   = \oint_{\gamma_2}\lambda.\cr}}
If a different choice is made for the $\gamma_i$, the pair $(a_D,a)$
will be transformed by an element of $SL(2,{\bf Z})$.

Now we would like to impose the condition that $\tau$ as defined
in \deftau\ has ${\rm Im}(\tau)>0$.  There is an obvious way to
satisfy this condition.  One has
\eqn\nurdef{\eqalign{{da_D\over du} & = \oint_{\gamma_1}{d\lambda\over
du}\cr
{da\over du} & = \oint_{\gamma_2}{d\lambda\over du}.\cr}}
Suppose that
\eqn\huxxo{{d\lambda\over du}= f(u) \lambda_1=f(u){dx\over y}}
with $f(u)$ a function of $u$ only.  Then we get
\eqn\weget{\eqalign{{da_D\over du}= & f(u)b_1\cr
                    {da\over du} =& f(u) b_2,\cr}}
with $b_i$ defined in \theper.
Consequently,
\eqn\eget{\tau={da_D/du\over da/du} = {b_1\over b_2}=\tau_u,}
and therefore ${\rm Im}(\tau)>0$.  Conversely, if ${\rm Im}(\tau)>0$
everywhere, then $\tau=\tau_u$ and $\lambda$ is as above.  To see this,
note first
if ${\rm Im}(\tau)>0$ everywhere, then for every $u$,
$\tau(u)$ is the $\tau$-parameter of an elliptic curve.
Also, the pair $(da_D/du,da/du)$ transforms under $SL(2,{\bf Z})$
the same way as $(a_D,a)$, and so gives a section of the same flat
bundle.    The family
of curves determined by $u\to\tau(u)$ consequently
has the same monodromies (and singularities) as
the family $E_u$ determined by $u\to \tau_u$.  It follows on general
grounds that these families coincide.

It remains to select $f$ and verify the desired asymptotic behavior of
$a_D$ and $a$ near $u=1,-1,\infty$.  We will in fact show that
everything works out correctly with $f=-\sqrt 2/4\pi$.
It will then be clear
that this is the unique choice that works; in particular, a non-constant
$f$ would somewhere introduce unwanted poles or zeroes.

First of all, given \huxxo, there is no problem in finding $\lambda$.
We can take
\eqn\wecan{\lambda = {\sqrt 2 \over 2\pi}
{dx\,\,\sqrt{x-u}\over \sqrt{x^2-1}}={ \sqrt 2
\over 2\pi}{dx\,\,y\over x^2-1}.}
Obviously, the $u$ derivative of this expression reproduces \huxxo\ with
$f=- \sqrt 2 /4\pi$.  The residues of $\lambda$ vanish since
\eqn\offor{\lambda={\sqrt 2 (\lambda_2 -u\lambda_1) \over 2\pi}}
with $\lambda_i$ as above.
$\lambda$ is odd under $P=w^2$, as the second formula in \wecan\ makes clear,
and that is why $P$ reverses the sign of $a$ and $a_D$ as defined presently.
\medskip
\fig{A non-trivial
one-cycle on $E_u$ comes from a contour in the
$x$ plane that loops around two of the branch points.
The cuts in the $x$ plane are indicated with dotted lines.}
{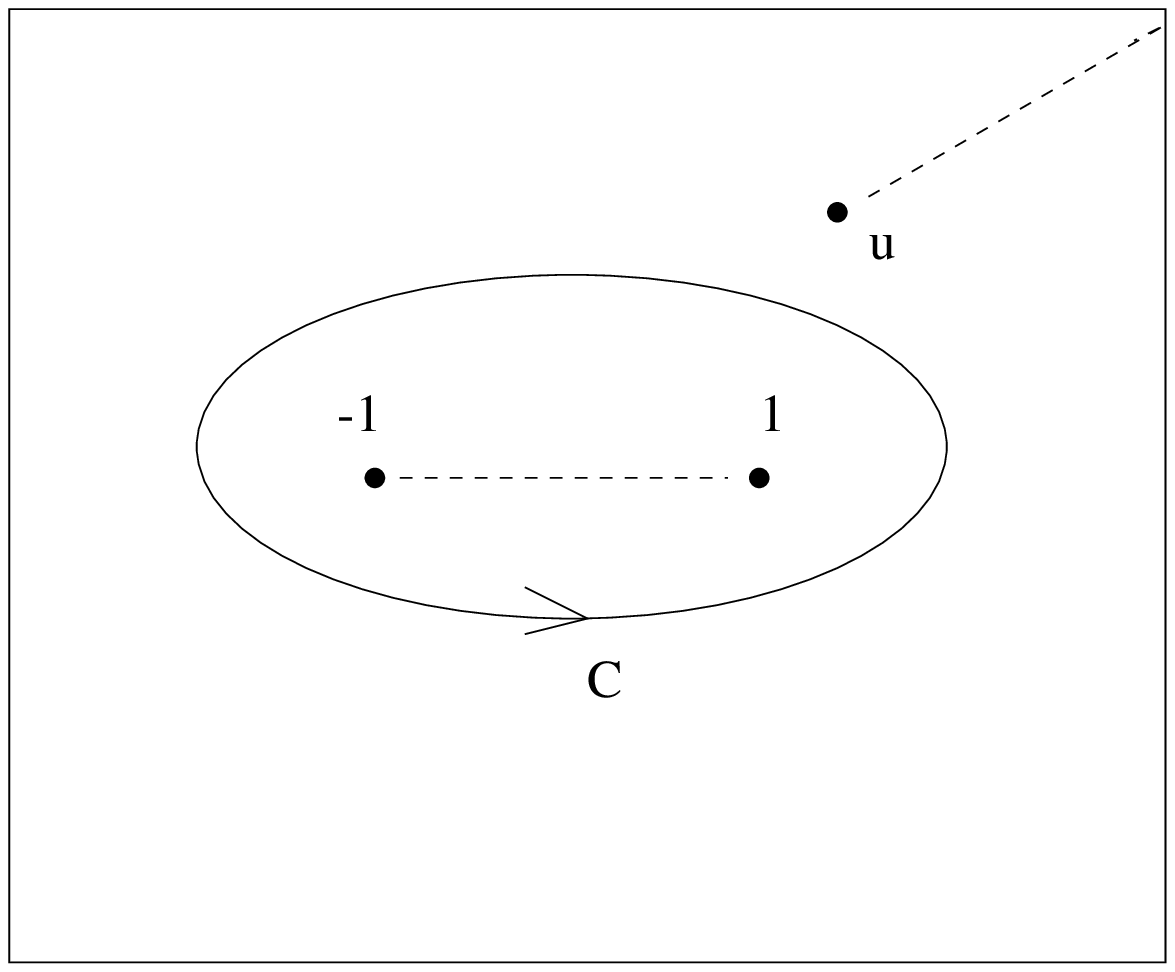}{10 truecm}
\figlabel\conto
\medskip

Now we need an explicit basis of one-cycles on $E_u$.
These can be constructed as follows.  A circle $C$ in the $x$ plane that
loops around two of the branch points lifts to a non-trivial cycle
$\gamma$ on $E_u$.  (Apart from any topological arguments, the
non-triviality of this cycle follows from the non-vanishing of the
integral that we are about to write.)  Suppose, for instance, that as in
 figure \conto\ we take $C$ to loop once around the branch points at
$x=1,-1$.  The differential form $\lambda$ that we wish to integrate,
viewed by the first formula in \wecan\ as a form on the $x$ plane, has a
factor $\sqrt{(x-u)/(x^2-1)}$ that requires branch cuts.  We
can take the cuts to run from $-1$ to $1$ and from $u$ to
$\infty$, as shown in figure \conto.
Then we can deform $C$ so it runs just around the cut between
$-1$ and $1$.    So
\eqn\wehav{\oint_\gamma\lambda={ \sqrt 2 \over 2\pi}
\oint_C {dx\,\sqrt{x-u}\over\sqrt{x^2-1}}
={ \sqrt 2 \over \pi}\int_{-1}^1{dx\,\sqrt{x-u}\over \sqrt{x^2-1}}.}
A factor of two in the last step comes because
the integral over $C$ contains an integral from
$-1$ to $1$ then from $1$ back to $-1$ on the other side of the cut;
the two segments make equal contributions.

Now we can implement the definition of $a$ and $a_D$ in \thendef.
Choosing the contour just described to be $\gamma_2$, we have
\eqn\whatsa{a={ \sqrt 2 \over \pi}\int_{-1}^1
{dx\,\sqrt{x-u}\over\sqrt{x^2-1}}.}
Similarly, defining another cycle $\gamma_1$
by using a circle that loops around
the branch points at $1$ and $u$, we
can take
\eqn\whatsad{a_D={ \sqrt 2
\over \pi}\int_1^u{dx\,\sqrt{x-u}\over\sqrt{x^2-1}}.}
Of course, this particular basis of one-cycles was picked with a view
to getting the desired behavior near the singularities in the $u$ plane.
The occurrence of a square root means that the overall signs of $a$ and
$a_D$ are ill-defined.  That is in keeping with the fact that, from the
outset, the classical relation $a^2=2 u$ means that $a$ can be
recovered from the gauge invariant quantity $u$ only up to sign.

It will suffice to study the behavior of $a$ and $a_D$ near $u=\infty$
and $u=1$, since the behavior near $u=-1$ is determined by the ${\bf Z}_2$
symmetry of the $u$ plane that was described earlier.

Near $u=\infty$, we get
\eqn\abeh{a\approx {\sqrt {2u} \over\pi}
\int_{-1}^1{dx\over\sqrt{1-x^2}}= \sqrt {2u}}
and (after a change of variables $x=uz$)
\eqn\adbeh{a_D={\sqrt {2u} \over\pi}\int_{1/u}^1{dz\,\sqrt{z-1}\over
\sqrt{z^2-u^{-2}}}.}
For $u\to\infty$ the integral develops a logarithmic divergence near
$z=0$; extracting the divergent term, we get
\eqn\nadbeh{a_D\approx i{\sqrt {2u} \ln u\over \pi}.}
Of course, this logarithm is the signal of asymptotic freedom
in the semiclassical region of large $u$.

The behavior near $u=1$ is equally easy to determine.  From \adbeh, we
get
\eqn\newad{a_D={\sqrt {2u} \over\pi}\int_{1/u}^1{dz \sqrt{z-1}
\over\sqrt{z+u^{-1}}\sqrt{z-u^{-1}}}\approx {1\over \pi} \int_{1/u}^1
{dz\,\sqrt{z-1}\over\sqrt{z-u^{-1}}}={i\over 2 }\left(1-{1\over u}
\right)\approx {i(u-1)\over 2 }.}
What about $a$?  At $u=1$ the integral for $a$ is:
\eqn\aval{a(u=1)={\sqrt 2 \over\pi}\int_{-1}^1{dx\over\sqrt{x+1}}={4
\over \pi} .}
However, the derivative of $a$ with respect to $u$ is given by an
integral
\eqn\ader{{da\over du}=-{\sqrt 2 \over 2\pi}\int_{-1}^1 {dx\over
\sqrt{(x+1)(x-1)(x-u)}} }
that -- for $u\to 1$ -- becomes logarithmically divergent
near $x=1$.  Extracting
the coefficient of the logarithm, we find that the expansion of $a$ is
\eqn\aexp{a={4\over\pi} -{(u-1)\ln(u-1)\over 2\pi }+ \dots.}
Comparing \aexp\ and \newad, we get the desired monodromy
$a\to a-2a_D$ near $u=1$.

This completes our verification of the expected properties.
But there is still one point to discuss.

{\it The Spectrum}

The remaining point concerns, in a sense, the physical meaning of
the duality that we have used to solve the theory.

In looping around $u=1$ or $u=-1$, the pair $(a_D,a)$ are transformed
by monodromies $M_1$ and $M_{-1}$.  The charges $(n_m,n_e)$ are
transformed similarly.  Naively, one would think that the spectrum
of BPS-saturated states would be transformed by the monodromy
matrices.  In that case, since the monodromies generate $\Gamma(2)$,
the spectrum of BPS-saturated states would be $\Gamma(2)$
invariant.

In fact, that is not true.  In the semiclassical region of large $u$,
the BPS-saturated states are the electrons and $W$ bosons
of $(n_m,n_e)=(0,\pm 1)$ and the dyons of $(n_m,n_e)=(\pm 1, n)$;
moreover the $W$ bosons are chiral multiplets (spin $\leq 1$)
and the dyons are hypermultiplets (spin $\leq 1/2$).  The fact
that this spectrum is not duality invariant is precisely the reason
that it was concluded many years ago that Olive-Montonen duality
did not hold for $N=2$ super Yang-Mills theory.

{}From the discussion at the end of section 4, there is a possible
mechanism for curing the problem.  The spectrum of BPS-saturated
states can jump on crossing a curve in the $u$ plane on which the ratio
$a_D/a$ is real.  If a curve on which $a_D/a$ is real passes through
$u=1$, then in looping around $u=1$, one would have to cross that curve
and the resulting jumping of the spectrum would invalidate conclusions
based on the monodromies.

It is easy to see that this situation does arise.  Near $u=1$,
$a$ is nearly real - in fact $a(u=1)=4 /\pi$. But  $a_D$ can have
an arbitrary phase near $u=1$ since $a_D\approx i(u-1)/2 $.
Thus, jumping can occur on a curve that near $u=1$
looks like $u=1+it,$ $t$ real, where $a_D/a$ is real.
A similar curve on which $a_D/a$
is real passes through $u=-1$.

It is consistent with everything we know, and will resolve
all the puzzles about lack of duality in the spectrum, if the
curve on which $a_D/a$ is real looks something like $|u|=1$.
Then one could avoid the jumping phenomenon only if one stays in the
region $u>1$; the only monodromy that can be seen in that region is
$M_\infty$, under which the spectrum of BPS-saturated states
is indeed invariant.  However, we do not know a practical way
to determine the curve on which $a_D/a$ is real.

There is, however, one important region in which one can easily prove
that $a_D/a$ is {\it not} real.  This is on the real $u$ axis for
$|u|>1$, as one can easily see from \whatsa\ and \whatsad.  For $u$ real
and $|u|>1$, the quantity $(x-u)/(x^2-1)$ is positive in the integration
region for $a$ and negative in the integration region for $a_D$, so $a$
is real and $a_D$ is imaginary.  In particular, by sticking to the real
$u$ axis, one can come in from the semiclassical region of large $u$ to
the singularities at $u=\pm 1$ without crossing any jumping curves.
Therefore, whatever particles become massless at $u=\pm 1$ must evolve
continuously from the BPS-saturated states that can be seen
semiclassically near infinity.  So for our picture of the strong
coupling region to make sense, the monopoles and dyons that we need for
the singularities must exist in the semiclassical region.  Happily, they
do.

\vskip 3cm
\centerline{{\bf Acknowledgements}}

We would like to thank T. Banks for a useful discussion on confinement
and condensation of magnetic monopoles, P. Deligne for some conceptual
assistance, G.  Moore for suggesting a possible role of
hypergeometric functions, and C. Vafa for explaining certain
two-dimensional results.  We have also benefitted from discussions with
K. Intriligator, R. Leigh, and S. Shenker.  This work was
supported in part by DOE grant \#DE-FG05-90ER40559 and in part by NSF
grant PHY92-45317.
\listrefs

\end